\documentclass[
10pt,aps,amsmath,amssymb,
reprint,
pra
,longbibliography
,notitlepage,
]{revtex4-1}
\usepackage{dcolumn}
\usepackage{bm}
\usepackage{braket}
\usepackage{xcolor}
\usepackage{graphicx}
\usepackage{mathtools}
\usepackage{multirow}
\usepackage[colorlinks=true,linkcolor=blue,urlcolor=black,citecolor=blue,bookmarksopen=true]{hyperref}

\begin{document}

\title{Optimizing Josephson-Ring-Modulator-based Josephson Parametric Amplifiers via full Hamiltonian control}
\author{Chenxu Liu}
\affiliation{Department of Physics and Astronomy, University of Pittsburgh}
\affiliation{Pittsburgh Quantum Institute}

\author{Tzu-Chiao Chien}
\affiliation{Department of Physics and Astronomy, University of Pittsburgh}
\affiliation{Pittsburgh Quantum Institute}

\author{Michael Hatridge}
\affiliation{Department of Physics and Astronomy, University of Pittsburgh}
\affiliation{Pittsburgh Quantum Institute}

\author{David Pekker}
\affiliation{Department of Physics and Astronomy, University of Pittsburgh}
\affiliation{Pittsburgh Quantum Institute}

\date{\today}

\begin{abstract}
    Josephson Parametric Amplifiers (JPA) are nonlinear devices that are used for quantum sensing and qubit readout in the microwave regime. While JPAs regularly operate near the quantum limit, their gain saturates for very small (few photon) input power. In a previous work, we showed that the saturation power of JPAs is not limited by pump depletion, but instead by the fourth-order nonlinearity of Josephson junctions, the nonlinear circuit elements that enables amplification in JPAs. Here, we present a systematic study of the nonlinearities in JPAs, we show which nonlinearities limit the saturation power, and present a strategy for optimizing the circuit parameters for achieving the best possible JPA. For concreteness, we focus on JPAs that are constructed around a Josephson Ring Modulator (JRM). We show that by tuning the external and shunt inductors, we should be able to take the best experimentally available JPAs and improve their saturation power by $\sim 15$~dB. Finally, we argue that our methods and qualitative results are applicable to a broad range of cavity based JPAs.  
\end{abstract}

\maketitle

\section{Introduction}\label{sec:intro}

Amplification is a key element in quantum sensing and quantum information processing. For example, readout of superconducting qubits requires a microwave amplifier that adds as little noise to the signal as possible~\cite{Clerk2010}, ideally approaching the quantum limit~\cite{Louisell1961, Gordon1963, Caves1982}. Recently, low-noise parametric amplifiers powered by the nonlinearity of Josephson junctions have been realized and are in regular use in superconducting quantum information experiments~\cite{Beltran2007, Castellanos-Beltran2008, Yamamoto2008,  Bergeal2010_exp, Bergeal2010_theory, Hatridge2011, Roch2012, Vijay2011, Johnson2012, Eichler2012}.

To evaluate the performance of a practical parametric amplifier there are three aspects that are equally important: (1) added noise at the quantum limit~\cite{Caves1982, Bergeal2010_exp, Bergeal2010_theory, Hatridge2011, Roch2012, Metelmann2014}, (2) broad-band amplification~\cite{Spietz2009, TC-Chien2019, Metelmann2014, Metelmann2015,zhong2019exceptional}, and (3) high saturation power~\cite{Kamal2009, Bergeal2010_theory, Abdo2013, Eichler2014, Kochetov2015, GQLiu2017, Frattini2018, Roy2018}, i.e. the ability to maintain desired gain for a large input signal power~\cite{Pozar2011}. The last requirement has been especially hard to achieve in Josephson parametric amplifiers and will be the focus of this paper. 

In previous works on Josephson parametric amplifiers, it was assumed that saturation power is limited by pump depletion~\cite{Kamal2009, Bergeal2010_theory, Abdo2013, Eichler2014, Roy2018}. This is a natural explanation, as the amplifier gain is a very sensitive function of the flux of the applied pump photons. Thus, as the input power is increased, and more pump photons are converted to signal photons, the gain falls. However, in Refs.~\cite{Kochetov2015, GQLiu2017, Frattini2018, Khan2019APS} it was pointed out that the fourth order nonlinear couplings (i.e. the Kerr terms), inherent in Josephson-junction based amplifiers, can also limit the saturation power. These terms induce a shift in the mode frequencies of the amplifier as a function of signal power, which can cause the amplifier to either decrease or increase its gain. Thus, we adopt the definition of saturation power as the lowest input power that causes the amplifier's gain to either increase or decrease by 1dB, which we abbreviate as $P_{\pm 1 \textrm{dB}}$. 

\begin{figure*}[htbp!]
    \centering
    \includegraphics[width = 6.5 in]{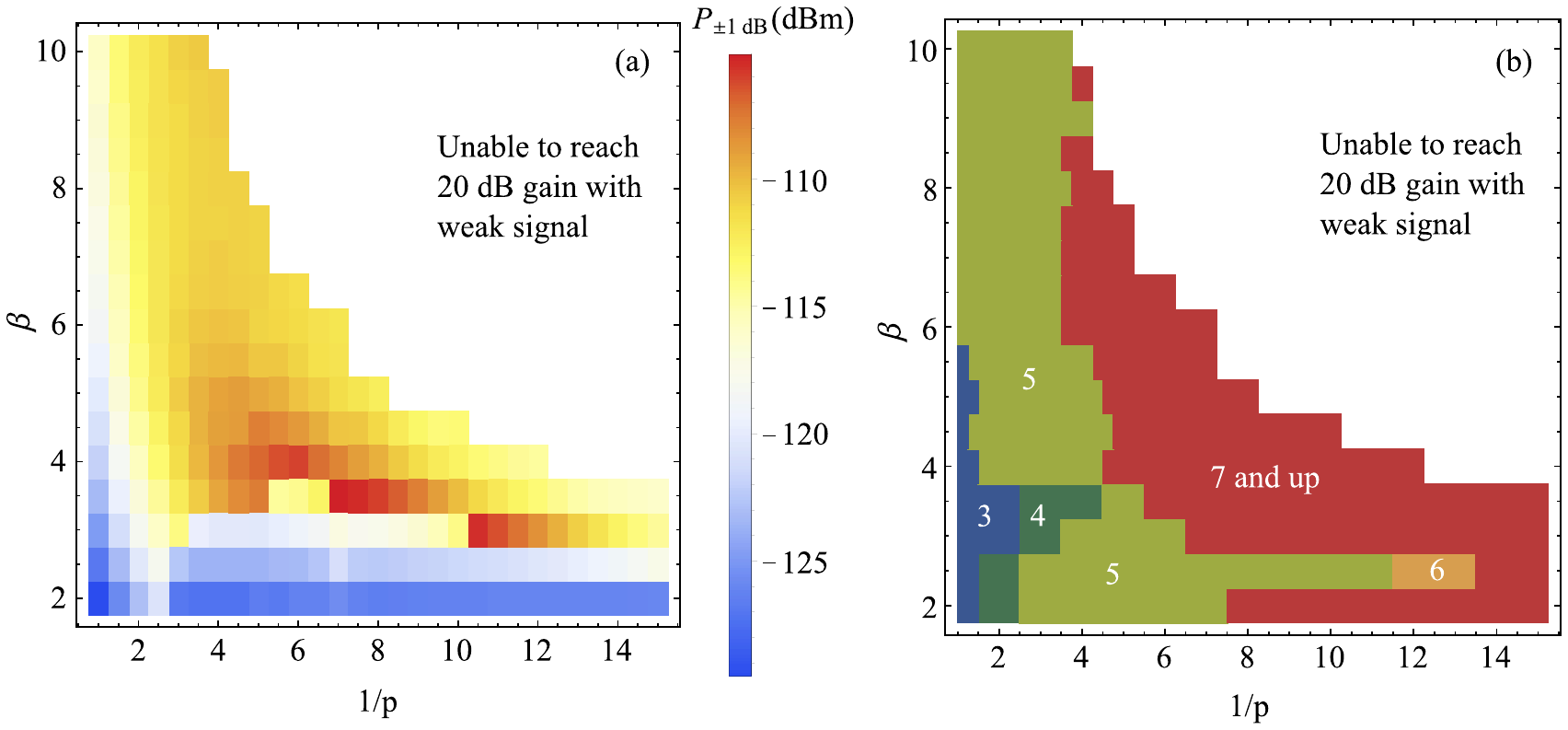}
    \caption{The saturation power for the JRM-based Josephson parametric amplifier with various JRM inductance ratio $\beta$ and participation ratio $p$ is shown in (a). 
    The amplifier has a sweet spot, in which saturation power is maximal, in the low $\beta$ and intermediate $p$ regime regime; Optimizing $\beta$ and $p$, we find saturation power of $P_{\pm 1~\textrm{dB}} \sim -104.8$~dBm at $\beta \approx 3.5$, $1/p \approx 7.0$.
    For small $p$ and large $\beta$, the amplifier is not able to reach the desired reflection gain (of $20$~dB); this region is labeled in white. In the intermediate $p$ regime ($1/p \sim 4$ to $10$), as we lower $\beta$, the saturation power first increases, hits the sweet spot, and then abruptly drops. To understand this behavior we refer to Fig.~\ref{fig:convergence}a, which shows that the gain at large signal powers tends to increase as $\beta$ decreases. This trend is at first beneficial to the amplifier, as the gain vs. signal power curve flattens out. However, at even lower $\beta$ the gain tends to increase with signal power (a feature that we call the ``shark-fin'') resulting in the amplifier saturating to $21$~dB (see Fig.~\ref{fig:convergence}a, $\beta=3$ curve) and hence the saturation power abruptly decreasing. The sweet spot of the saturation power is located at the edge of the this ``reflection gain boost'' regime. In (b), we show the minimum truncation order needed to converge small-signal reflection gain of the amplifier to $20 \pm 0.3$~dB. In our main text, we show that the convergence order of small-signal reflection gain gives a good prediction on the convergence order of the saturation power (see Fig.~\ref{fig:convergence}b \& c). In the small-$\beta$, large-$p$ corner, the third order truncation is enough to make the time-solver convergence to the desired $20$~dB reflection gain. While as we decrease the participation ratio, the higher and higher order is needed to converge the truncated theory, which shows that the full-order simulation is needed to predict the performance of the amplifier near the sweet spot.}
    \label{fig:result}
\end{figure*}

In this paper, we address the question: for a given device, does pump depletion, Kerr terms, or higher-order nonlinearities limit the saturation power $P_{\pm 1 \textrm{dB}}$? Yet how do we tame these limitations to optimize the device by maximizing $P_{\pm 1 \textrm{dB}}$? Our analysis and results are generally applicable for all amplifiers based on third-order couplings, including JPAs based on Superconducting Nonlinear Asymmetric  Inductive eLements (SNAILs)~\cite{Frattini2017,Frattini2018, Sivak2019, sivak2019arXiv}), flux pumped Superconducting QUantum Interference Devices (SQUIDs)~\cite{Yamamoto2008, Aumentado2010, Mutus2013, Zhou2014, naaman2017}, and the Josephson Parametric Converters (JPCs)~\cite{Bergeal2010_exp, Bergeal2010_theory, GQLiu2017, Abdo2011, Abdo2013}). These techniques we develop may also be of use in the simulation of non-cavity based amplifiers, such as the traveling wave parametric amplifier (TWPA)~\cite{Macklin2015, White2015, Zorin2016}.

\begin{figure*}[t]
    \centering
    \includegraphics[width = 6.0 in]{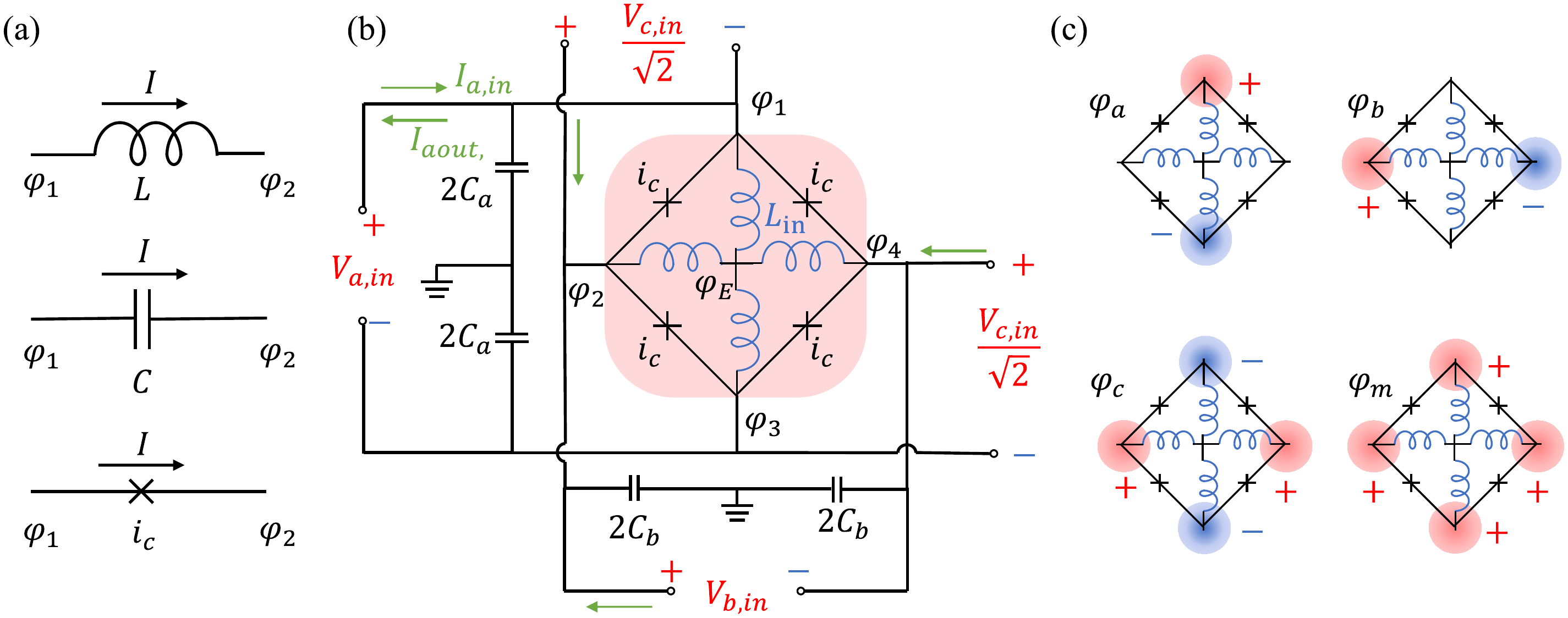}
    \caption{In (a) we show the typical circuit elements that we will focus on in this paper, a linear inductor with inductance $L$, a capacitor with capacitance $C$ and a Josephson junction with critical current $i_c$. The node phase and the convention for the current is labeled on each element drawings. The circuit model for the JRM-based JPA circuit is shown in (b). The circuit model for a linear inductance shunted Josephson Ring Modulator (JRM) is shaded in red. We connected the linear inductance shunted JRM with the capacitors and the input-output ports. We assume the normal modes are symmetrically driven by the ports. For the port corresponds to mode $\varphi_a$, we use the green arrows to show the input and output (reflected) current flow direction. For $b$-mode, we only use the green arrow to show the flow direction of the input current. The connection of the $c$-port is not shown in the plot. The $c$-port also drives the corresponding mode profile symmetrically. The corresponding normal modes, including three nontrivial modes $\varphi_a$, $\varphi_b$ and $\varphi_c$ and one trivial mode $\varphi_m$, are shown in (c).}
    \label{fig:JRM_modes}
\end{figure*}

In the JPC, three microwave modes (a,b,c) are coupled via a ring of four Josephson Junctions [the so-called Josephson Ring Modulator (JRM), see Fig.~\ref{fig:JRM_modes}(b) shaded part, for example]. A third-order coupling ($g_3 \varphi_a \varphi_b \varphi_c$) between the fluxes ($\varphi_i$) of three microwave modes is obtained by applying a static magnetic flux to the JRM ring.  Phase-preserving gain is obtained by pumping one mode (typically c) far off resonance at the sum frequency of the other two (a and b), with the gain amplitude being controlled by the strength of the pump drive.  

We now discuss the main results of our investigation, which are summarized in Fig.~\ref{fig:result}. Previously, descriptions of JPC's relied on expanding the nonlinear couplings between the three microwave modes in a power series of cross- and self- couplings. The power series was truncated at the lowest possible order, typically fourth (i.e. corresponding to the cross- and self-Kerr terms)~\cite{Bergeal2010_theory,Roch2012, GQLiu2017,TC-Chien2019}. In the present paper, we compare these power series expansions with the exact numerical solutions in the framework of semi-classic input-output theory. Our first main finding is that there is indeed a sweet spot for operating a JPA, see Fig.~\ref{fig:result}(a), at which $P_{\pm 1 \textrm{dB}}$ is maximized. The sweet spot appears for moderate values of the two circuit parameters: participation ratios $p \sim 1/7$ and shunt inductance ($\beta=L_\text{J}/L_{\text{in}} \sim 3.5$, where $L_{\text{in}}$ is the shunt inductance, $L_\text{J}= \varphi_0/I_0$ is the Josephson inductance, $\varphi_0=\hbar / 2 e$ is the reduced flux quantum, and $I_0$ is the Josephson junction critical current). Our second main finding is that in the vicinity of the sweet spot nonlinear terms up to at least 7th order are comparable in magnitude and hence truncating the power series description at fourth order is invalid, see Fig.~\ref{fig:result}(b). The second main result can be interpreted from two complementary perspectives. First, the sweet spot corresponds to high pump powers and hence the energy of Josephson junctions cannot be modeled by a harmonic potential anymore. Second, different orders of the power series expansion have either a positive or a negative effect on the gain as a function of signal power; when the magnitudes of terms at different orders are comparable the terms cancel each other resulting in a boost of $P_{\pm 1 \textrm{dB}}$. We hypothesize that the second main finding is a generic feature for Josephson junction based parametric amplifiers.

Before moving to a detailed development of our theory, we provide a summary of the key steps of our investigation and outline the structure of our paper.

We begin by noting that in addition to the above-mentioned parameters $p$ and $\beta$, the magnetic flux through the JRM $\varphi_{\text{ext}}=(2 \pi/\varphi_0) \Phi_{\text{ext}}$ is another important control parameter. For conventional JRMs~\cite{Bergeal2010_theory, Bergeal2010_exp}, at non-zero values of applied flux there are non-zero cross- and self-coupling at all orders (4th, 5th, etc.). However, we have recently realized that a linearly-shunted variant of the JRM~\cite{Roch2012, TC-Chien2019} can null all even-order couplings at a special flux bias point ($\varphi_{\text{ext}}=2\pi$), which we call the Kerr nulling point. The same nulling is also observed in SNAIL-based devices~\cite{Frattini2017}. In the context of a JPC with participation ratio $p<1$, even couplings come back but remain much smaller than at generic values of $\varphi_{\text{ext}}$. Therefore throughout this paper, we focus on $\varphi_{\text{ext}}$ at or in the vicinity of the Kerr nulling point.

We calculate the saturation power using semi-classical equations of motion for the microwave modes, which are derived using input-output theory from the Lagrangian for a lumped-circuit model of the JPA. When we consider higher than third-order couplings, these equations are not generally analytically solvable. To analyze the saturation power for a given set of parameters, we compare numerical integration of the full nonlinear equations to solutions of various, artificially truncated versions of the equations obtained using both numerical integration and perturbation theory. 
We begin by investigating the effects of pump depletion. To do so, we analyze the dynamics of all the modes with interactions truncated at third order. Using classical perturbation theory to eliminate the dynamics of the pump mode ($c$), we find, in contradiction with the basic understanding of pump `depletion', that the first corrections are a complex fourth order cross-Kerr coupling between modes $a$ and $b$, and an associated two-photon loss process in which pairs of $a$ and $b$ photons decay into the $c$ mode, that effectively \textit{increase} the pump strength.  The dynamically generate Kerr terms act similarly to the intrinsic Kerr terms, including giving rise to saturation to higher gain when the pump mode frequency is positively detuned from the sum frequency.  Further, in the shunted JRM, we can partially cancel the real part of the dynamically generated Kerr by tuning the applied flux near the Kerr nulling-point so as to generate an opposite sign intrinsic Kerr.  Thus, the presence of judicious intrinsic Kerr can be a virtue, and the ultimate pump `depletion' limit is set by the imaginary Kerr and two-photon loss.  Increasing the $\beta$ value of the JRM reduces these effects and increase the JPCs saturation power.  Away from the nulling point, these depletion effects are overwhelmed by the intrinsic Kerr effects, and the device is Kerr-limited in agreement with our previous results.
  
Next, we perform calculations with full nonlinearity, and find that saturation power stops increasing at high $\beta$.  We find that this is primarily due to certain 5th order terms of the form $\left( \varphi_a^2 + \varphi_b^2 \right) \varphi_a \varphi_b \varphi_c$. These terms modulate the effective parametric coupling strength as a function of the input signal power thus shifting the amplifier away from the desired gain by increasing the effective parametric coupling (in fact, throughout this work we failed to identify a scenario in which the amplifier `runs out of pump power').  

To suppress the strength of these terms relative to the desired third order coupling, we introduce an additional control knob by adding outer linear inductors $L_\text{out}$ in series with the JRM. The participation ratio $p=L_{\text{JRM}}/(2 L_\text{out}+L_{\text{JRM}})$, where $L_{\text{JRM}}$ is the effective inductance of the JRM, controls what fraction of the mode power is carried by the JRM. Decreasing $p$ results in the suppression of all coupling terms; however, the higher-order coupling terms decrease faster than the lower order ones. Thus, if the saturation power is limited by intrinsic 5th order terms, we can increase the saturation power by decreasing the participation ratio $p$. We remark that as the pump power is increased, the cross-coupling terms result in a shift of the JPA frequencies that must be compensated, which we do for each value of $p$ and $\beta$. Tuning both $p$ and $\beta$ we can find a sweet spot for the operation of the JPC, as discussed above.

In general, the mode frequencies shift with applied pump power.  This, combined with the fact that JPAs can function with pump detunings comparable to the bandwidth of the resonators on which they are based, makes comparing theory and experiment very complicated.  For concreteness, our simulations vary the applied pump and signal frequencies to identify the bias condition which requires minimum applied pump power to achieve 20 dB of gain. These points can be readily identified in experiment~\cite{GQLiu2017}.  However, there has been a recent observation in SNAIL-based JPAs that deliberate pump detuning can additionally enhance device performance~\cite{Sivak2019}, and serve as an {\it in situ} control to complement the Hamiltonian engineering we discuss in this work. 

This paper is organized as follows. In Sec.~\ref{sec:base_theory}, we focus on the closed model of JPA circuit (without input-output ports). We start by reviewing the basic theory of circuits with inductors, capacitors and Josephson junctions in subsection~\ref{subsec:circuit_element}. In subsection~\ref{subsec:normal_mode}, we include the external shunted capacitors with JRM, and present the normal modes of the JPA circuit model using Lagrangian dynamics. In Section~\ref{sec:nonlinear}, we further include the input-output ports into the circuit model of the JPA, and construct the equations of motion to describe the dynamics of the circuit. In Section~\ref{sec:optimization}, we investigate the limitation on the saturation power of the JPA without external series inductors. Specifically, we analyze the 3rd order theory using both numerical and perturbative approaches in Sec.~\ref{subsec:SoP_3rd}. We compare these results with the effect of Kerr nonlinearities in Sec.~\ref{subsec:StP_4th} and identify the dynamically generated Kerr terms and the two-photon loss processes. Intrinsic fifth- and higher-order nonlinear couplings are investigated in Sec.~\ref{subsec:higher_order}. We put these results together in Sec.~\ref{subsec:full_order} and identify which effect is responsible for limiting the saturation power in different parametric regimes. In Sec.~\ref{sec:discussion} we consider the consequence of the series inductors outside of it. We show that the series inductors, which suppress the participation ratio of the JRM, can be used to improve the dynamic range of the JRM. We discuss how to optimize the saturation power of the JPA, taking into account both series inductors and full nonlinearities in Section~\ref{sec:full_optimization}. In Sec.~\ref{sec:exp_imperfect}, we further explore how the saturation power is affected by the magnetic field bias, the modes' decay rates and stray inductors in series of the Josephson junction in JRM loop. We provide an outlook on the performance of Josephson junction based amplifiers in Section~\ref{sec:summary}.

\section{Equations of motion for circuits made of inductors, capacitors, and Josephson junctions} \label{sec:base_theory}

In this section, we review the theory of lumped circuits elements. We start from the Lagrangian treatment of single circuit elements in subsection~\ref{subsec:circuit_element}. Then in subsection~\ref{subsec:normal_mode}, we work on the JRM and the closed JPA circuit model and solved the normal mode profiles of the JRM.


\subsection{Lagrangian description of linear inductance, Josephson junctions and capacitors} \label{subsec:circuit_element}

The equations of motion (EOM) that describe the dynamics of a circuit with Josephson junctions, inductors, and capacitors can be derived using the formalism of Lagrangian dynamics, which naturally leads to Kirchhoff's law. We use the dimensionless flux on each node of the circuit, $\varphi_j(t) = \frac{1}{\phi_0} \int_{-\infty}^{t} V_j (t') dt'$, as the set of generalized coordinates. The Lagrangian $\mathcal{L}[\{\varphi_j,\dot{\varphi_j}\}]$ is defined as
\begin{align}
    \mathcal{L}=T[\{\dot{\varphi_j}\}]-U[\{\varphi_j\}],
\end{align}
where $T$ is the kinetic energy associated with the capacitors and $U$ is the potential energy associated with the inductors and the Josephson junctions. Using Fig.~\ref{fig:JRM_modes}(a) to define the nodes and current direction for each type of circuit element, we observe that each capacitor contributes
\begin{align}
    E_C & = \frac{C}{2} \phi_0^2 \left( \dot{\varphi}_1-\dot{\varphi}_2 \right)^2 \label{eq:capacitor_energy}
\end{align}
to $T[\{\dot{\varphi_j}\}]$, while each inductor and each Josephson junction contributes
\begin{align}
     E_L & = \frac{\phi_0^2}{2 L} (\varphi_2 - \varphi_1)^2,\\
     E_\text{J} & = - \phi_0 i_c \cos\left( \varphi_2 - \varphi_1 \right),
\end{align}
to $U[\{\varphi_j\}]$, where $i_c$ is the critical current of the Josephson junctions. The current across a capacitor is $-\frac{1}{\phi_0} (\delta E_C/\delta \varphi_1) =  \frac{1}{\phi_0} (\delta E_C/\delta \varphi_2)$, while the current across an inductor is $\frac{1}{\phi_0} (\delta E_L/\delta \varphi_1) = - \frac{1}{\phi_0} (\delta E_L/\delta \varphi_2)$ and across a Josephson junction $\frac{1}{\phi_0} (\delta E_\text{J}/\delta \varphi_1) = - \frac{1}{\phi_0} (\delta E_\text{J}/\delta \varphi_2)$. Using the Lagriangian $\mathcal{L}$ of the circuit elements, the current that flows out of each node of the circuit is $J_j= - \frac{1}{\phi_0} (\delta \mathcal{L}/\delta \varphi_j)$. To obtain the equations of motion (EOMs) we extremize the action by setting $J_j=0$, which corresponds to enforcing Kirchhoff's law.

Next, we apply the Lagrangian formalism to derive the potential energy of the linear-inductor-shunted JRM, the key component at the heart of the JPA, shown in Fig.~\ref{fig:JRM_modes}(b), which is shaded in red. The potential energy of the JRM circuit~\footnote{Throughout our paper, we make the assumption that the JRM circuit is symmetric, that is all four inner inductors are identical and all four Josephson junctions are identical.} is
\begin{align}
    E_{\textrm{JRM}} = & \frac{\phi_0^2}{2 L_\textrm{in}}\sum_{j=1}^4 \left( \varphi_j - \varphi_{E} \right)^2 \nonumber \\
    & - \phi_0 i_{c} \sum_{i=1}^4 \cos\left[ \varphi_{i+1}-\varphi_i - \frac{\varphi_\textrm{ext}}{4} \right],
\label{eq:raw_HJRM}
    \end{align}
where $\varphi_j$'s are the phases of the superconductors on the  nodes [see Fig.~\ref{fig:JRM_modes}(b)] and we adapt the convention that $\varphi_5 = \varphi_1$ for the summation. The external magnetic flux though the JRM circuit $\Phi_{\textrm{ext}}$ controls the parameter $\varphi_{\textrm{ext}} = \Phi_{\textrm{ext}} / \phi_0 $. Applying Kirchhoff's law to node $E$, we obtain $\varphi_E = \frac{1}{4}\left(\varphi_1 + \varphi_2 + \varphi_3 + \varphi_4 \right)$.

\subsection{Normal modes of the Josephson Parametric Amplifier} \label{subsec:normal_mode}

In this sebsection, we focus on the equations of motion of the closed circuit model of the JPA (i.e. ignore the input-output ports) and analyze the normal mode profile of the JPA circuit (see Fig.~\ref{fig:JRM_modes}(b), but without input ports).

The potential energy of the shunted JRM was derived in the previous subsection, see Eq.~\eqref{eq:raw_HJRM}. The kinetic energy associated with the capacitors [Fig.~\ref{fig:JRM_modes}(b)], Eq.~\eqref{eq:capacitor_energy}, is
\begin{equation}
    E_c = \phi_0^2 \left( C_a \dot{\varphi_1}^2 +  C_b \dot{\varphi_2}^2 +  C_a \dot{\varphi_3}^2+ C_b \dot{\varphi_4}^2\right),
\end{equation}
Which gives the Lagrangian $\mathcal{L} = E_c - E_{\textrm{JRM}}$. The EOM of this closed circuit can be constructed using Lagrange's equation, e.g. for a node flux $\varphi_j$,
\begin{align}
2 C_j \ddot{\varphi_j} & + \frac{1}{L_\textrm{in}} \left(\varphi_j - \varphi_E\right) + \frac{1}{L_\text{J}}\left[\sin\left(\varphi_j - \varphi_{j+1} + \frac{\varphi_\textrm{ext}}{4}\right)\right. \nonumber \\
& \left. - \sin\left( \varphi_{j-1} - \varphi_j + \frac{\varphi_\textrm{ext}}{4}\right)\right] = 0,
\label{eq:close_node1}
\end{align}
where $\varphi_j$ is the node phases, $j = 1, 2, 3, 4$, and we use the index convention that $\varphi_0 = \varphi_4$, $\varphi_5 = \varphi_1$. According to the Fig.~\ref{fig:JRM_modes}(b), the node capacitance are $C_1 = C_3 = C_a$ and $C_2 = C_4 = C_b$. The Josephson inductance $L_\textrm{J} = \phi_0 / i_c$.


To analyze the normal modes of the circuit, we assume we have chosen suitable values for the parameters so that the ground state of the circuit is $\varphi_1 = \varphi_2 = \varphi_3 = \varphi_4 = 0$, and expand in small oscillations to obtain a linearized set of EOMs around the ground state. 
The corresponding normal coordinates, which we denote as $[\varphi_M]$ in vector form, are related to the node fluxes via $[\varphi] = [\mathcal{A}].[\varphi_M]$, where transformation matrix $[\mathcal{A}]$ is 
\begin{equation}
    [\mathcal{A}] = \left(
    \begin{array}{cccc}
        1   & \frac{1}{2}   & 0             & -\frac{C_b}{\left(C_a+C_b\right)}    \\
        1   & 0             & \frac{1}{2}   & \frac{C_a}{\left(C_a+C_b\right)}           \\
        1   & -\frac{1}{2}  & 0             & -\frac{C_b}{\left(C_a+C_b\right)}    \\
        1   & 0             & -\frac{1}{2}  & \frac{C_a}{\left(C_a+C_b\right)}           \\
    \end{array}
    \right), 
    \label{eq:model_matrix}
\end{equation}
and the flux coordinates vectors are defined as $[\varphi] = \left( \varphi_1, \varphi_2, \varphi_3, \varphi_4\right)^\intercal$ and $[\varphi_M] = \left( \varphi_m, \varphi_a, \varphi_b, \varphi_c\right)^\intercal.$
Inverting this transformation, we obtain the expression for the normal modes in terms of the node fluxes,
\begin{subequations}
\label{eq:mode_raw}
\begin{align}
        \varphi_a  & = \varphi_1 - \varphi_3, \quad \\
        \varphi_b  & = \varphi_2 - \varphi_4, \quad \\
        \varphi_c  & = -\frac{1}{2} \left( \varphi_1 + \varphi_3 - \varphi_2 - \varphi_4 \right) \\
        \varphi_m & = \frac{C_a}{2(C_a+C_b)}\left( \varphi_1 + \frac{C_b}{C_a}\varphi_2 + \varphi_3 + \frac{C_b}{C_a}\varphi_4 \right).
\end{align}
\end{subequations}
The profiles for the normal modes, $\varphi_a$, $\varphi_b$, $\varphi_c$ and $\varphi_m$ are sketched in Fig.~\ref{fig:JRM_modes}(c). The normal mode $\varphi_m$ has zero frequency and it is not coupled with any of the other three modes [see Eq.~\eqref{eq:HJRM}]. Therefore, $\varphi_m$ is a trivial mode, which can be safely ignored in our following discussion. The corresponding frequencies for the other three nontrivial modes are
\begin{subequations}
\label{eq:mode_f}
\begin{align}
\omega_a^2 & = \frac{L_\text{J}+ 2 L_\textrm{in}\cos\left(\frac{\varphi_\textrm{ext}}{4}\right)}{2 C_a L_\textrm{in}L_\text{J}}, \\
\omega_b^2 &= \frac{L_\text{J}+ 2 L_\textrm{in}\cos\left(\frac{\varphi_\textrm{ext}}{4}\right)}{2 C_b L_\textrm{in}L_\text{J}},  \\
\omega_c^2 &= \frac{C_a+C_b}{C_a C_b} \cdot\frac{L_\text{J}+ 4 L_\textrm{in}\cos\left(\frac{\varphi_\textrm{ext}}{4}\right)}{4 L_\textrm{in}L_\text{J}}.
\end{align}
\end{subequations}

With the coordinate transformation given by the model matrix $[\mathcal{A}]$ [see Eq.~\eqref{eq:model_matrix}], we can re-write the potential energy of the JPA (the energy of JRM circuit) using the normal modes $\varphi_a$, $\varphi_b$ and $\varphi_c$, as
\begin{widetext}
\begin{equation}
    \begin{aligned}
        E_{\textrm{JRM}} = & -4 E_{\textrm{J}} \left[ 
            \cos\left(\frac{\varphi_a}{2} \right) \cos\left(\frac{\varphi_b}{2}\right) \cos\left(\varphi_c\right) \cos\left( \frac{\varphi_\textrm{ext}}{4}\right)
            + \sin\left(\frac{\varphi_a}{2} \right) \sin\left(\frac{\varphi_b}{2}\right) \sin\left(\varphi_c\right) \sin\left( \frac{\varphi_\textrm{ext}}{4}\right)
            \right] \\
            & +\frac{\phi_0^2}{4 L_\textrm{in}} \left(\varphi_a^2 + \varphi_b^2 + 2 \varphi_c^2\right)
    \end{aligned}
\label{eq:HJRM}
\end{equation}
where $E_{\textrm{J}} = \phi_0 i_c$ is the Josephson energy. 

We observe from Eq.~\eqref{eq:HJRM} that the four Josephson junctions on the outer arms of the JRM provide nonlinear couplings between the normal modes of the circuit. Assuming that the ground state of the circuit is $\varphi_a = \varphi_b = \varphi_c = 0$, and it is stable as we tune the external magnetic flux bias, we can expand the nonlinear coupling terms around the ground state as
\begin{equation}
    \begin{aligned}
    E_{\textrm{JRM}} \sim & 
        \left[ 
            \frac{\phi_0^2}{4L_\textrm{in}} + \frac{E_\text{J}}{2} \cos
                \left( 
                    \frac{\varphi_{\textrm{ext}}}{4}
                \right)
        \right]
        \left( 
            \varphi_a^2 + \varphi_b^2
        \right) + 
        \left[ 
            \frac{\phi_0^2}{2L_\textrm{in}} + 2 E_\text{J}  \cos
                \left( 
                    \frac{\varphi_{\textrm{ext}}}{4}
                \right)
        \right] \varphi_c^2 
     - E_\text{J} \sin
        \left( 
            \frac{\varphi_\textrm{ext}}{4}
        \right) \varphi_a \varphi_b \varphi_c \\
    & - \frac{1}{96} E_\text{J} \cos
        \left( 
            \frac{\varphi_\textrm{ext}}{4}
        \right) 
        \left( 
            \varphi_a^4 + \varphi_b^4 + 16 \varphi_c^4
        \right) 
     + \frac{1}{16} E_\text{J} \cos
        \left( 
            \frac{\varphi_\textrm{ext}}{4} 
        \right) 
        \left( 
            \varphi_a^2 \varphi_b^2 + 4 \varphi_a^2 \varphi_c^2 + 4 \varphi_b^2 \varphi_c^2 
        \right) + ...
    \end{aligned}
    \label{eq:H_expansion}
\end{equation}
\end{widetext}
Because of the parity of the cosine and sine functions, the cosine terms in Eq.~\eqref{eq:HJRM} contribute the even order coupling terms while the sine terms contributes the odd order coupling terms. The nonlinear couplings are controlled by the external magnetic flux bias $\varphi_\textrm{ext}$. The third order nonlinear coupling is the desired term for a non-degenerate Josephson parametric amplifier, while all the higher order couplings are unwanted. The Kerr nulling point~\cite{Frattini2017,TC-Chien2019} is achieved by setting the external magnetic flux to $\varphi_{\textrm{ext}} = 2 \pi$ (and assuming that the ground state $\varphi_a=\varphi_b=\varphi_c=0$ remains stable), and we find that all the even order nonlinear couplings are turned off.

\section{Input-output theory of the Josephson Parametric Amplifier} \label{sec:nonlinear}

The linear-inductor shunted JRM described in the previous section is the core elements of the Josephson parametric amplifier. In order to build the JPA, we add input-output lines and external parallel capacitors to the JRM, see Fig.~\ref{fig:JRM_modes}(b). In Section~\ref{sec:discussion} we will extend the description of the JPA by adding stray and series inductors to the JRM. 

In order to fully model the JPA, we need to describe the input-output properties of the JPA circuit. In Subsec.~\ref{subsec:IO_theory} we introduce input-output theory, and apply it to the problem of modeling drive and response of the JPA. In Subsec.~\ref{subsec:full_order_EOM} we present the full nonlinear equations of motion that describe the JPA circuit.

\subsection{Input-output relation for the Josephson Parametric Amplifier} \label{subsec:IO_theory}
To solve the full dynamics of the JPA with amplification process, we need to be able to describe the microwave signals that are sent into and extracted (either reflected or transmitted) from the circuit. Therefore, we need to connect the input-output ports to the JPA circuit and include the description of them in the EOMs. 

To simplify the problem, we assume that the drives perfectly match the profiles of the corresponding normal modes, as shown schematically for modes $a$ and $b$ in Fig.~\ref{fig:JRM_modes}(a). Take mode $a$ as an example. We send in a microwave signal with the amplitude of the voltage $V_{a,\textrm{in}} = \phi_0 \dot{\varphi}_{a,\textrm{in}}$ into the port for this mode. The corresponding current flow from the transmission line to the amplifier is $I_{a,in} = \frac{V_{a,in}}{Z_a}$, where $Z_a$ is the impedance of the transmission line. The voltage applied to node $1$ and node $3$ are $V_1 = \frac{\phi_0}{2} \dot{\varphi}_{a,\textrm{in}}$ and $V_3=-\frac{\phi_0}{2} \dot{\varphi}_{a,\textrm{in}}$, respectively. While the output microwave signal has output voltage amplitude $V_{a,\textrm{out}} = \phi_0 \dot{\varphi}_{a,\textrm{out}}$ and the output current is $I_{a,\textrm{out}}=\frac{V_{a,\textrm{out}}}{Z_a}$. 

At the nodes which connect to the transmission line, e.g. nodes $1$ and $3$ for $a$ mode, the voltage and current should be single-valued. This requirement induces an input-output condition
\begin{equation}
    \begin{aligned}
        V_{a,\textrm{in}} + V_{a,\textrm{out}} & = V_a = V_1 - V_3 \\
        I_{a,\textrm{in}} - I_{a,\textrm{out}} & = I_{1,a} = -I_{3,a},
    \end{aligned}
\end{equation}
where $I_{1,a}$ ($I_{3,a}$) is the net current flow into node $1$ ($3$) of the amplifier from the port. Because the output signals should be determined by the input signals, we eliminate the output variables from the input-output relation so that it can be combined with the current relation inside the JRM to construct the EOMs for the open circuit model
\begin{equation}
    I_{1,a} = - I_{3,a} = \frac{2 V_{a,\text{in}}}{Z_a} - \frac{\phi_0 \left( \dot{\varphi_1} - \dot{\varphi_3} \right)}{Z_a}.
\label{eq:IO_aI}
\end{equation}
Given the the drives (inputs), we can solve for the mode fluxes using the EOMs, and then obtain the outputs using the input-output relations. For example, the output voltage on port $a$ is determined by
\begin{equation}
    V_{a,\textrm{out}} = \phi_0 \left( \dot{\varphi}_1 - \dot{\varphi}_3 \right) - V_{a,\textrm{in}}.
\label{eq:IO_aV}
\end{equation}

In the remainder of this paper we focus on the reflection gain of the JPA which is obtained from a phase-preserving amplification process. The input signal to be amplified by the JPA is a single-frequency tone. The amplified output is the reflected signal at the same frequency. Using the Josephson relation relating voltage and flux, we observe that the reflected voltage gain is equal to the reflected flux gain. Therefore, we use the input-output relation for the mode flux, e.g. for port $a$ we have
\begin{equation}
    \varphi_{a,\textrm{out}} = \varphi_1 - \varphi_3 - \varphi_{a,\textrm{in}}.
    \label{eq:IO_aPhi}
\end{equation}

The analysis of input-output ports for mode $b$ and $c$ is similar. For $b$-port we have
\begin{subequations}
    \label{eq:IO_b}
    \begin{align}
        I_{2,b} & = -I_{4,b} = \frac{2 V_{b,\textrm{in}}}{Z_b} - \frac{\phi_0 \left( \dot{\varphi_2} - \dot{\varphi_4} \right)}{Z_b}  \label{eq:IO_b_I}\\
        V_{b,\textrm{out}} & = \phi_0 \left( \dot{\varphi}_2 - \dot{\varphi}_4 \right) - V_{b,\textrm{in}}, \label{eq:IO_b_V}
    \end{align}
\end{subequations}
and for $c$-port
\begin{subequations}
    \label{eq:IO_c}
    \begin{align}
        I_{2,c} & = I_{4,c} = - I_{1,c} = -I_{3,c} \nonumber \\
        & = \frac{\sqrt{2} V_{c,\textrm{in}}}{Z_c} - \frac{\phi_0}{2 Z_c}\left( \dot{\varphi}_2 + \dot{\varphi}_4 - \dot{\varphi}_1 - \dot{\varphi}_3\right) \label{eq:IO_c_I}\\
        V_{c,\textrm{out}} & = \frac{\sqrt{2}\phi_0}{2} \left( \dot{\varphi}_2 + \dot{\varphi}_4 - \dot{\varphi}_1 - \dot{\varphi}_3\right) - V_{c,\textrm{in}}. \label{eq:IO_c_V}
    \end{align}
\end{subequations}
The extra factor $\sqrt{2}$ that app[ears for the $c$ port is due to the microwave power being split $50/50$ between the two transmission lines that drive all four nodes simultaneously.

When constructing the EOM with input-output ports, we should consider the current contribution from all the input-output ports together. For example, the net current injected through node $1$ should have contributions from the drive applied to both ports for mode $a$ and mode $c$, i.e. $I_{1,\textrm{net}} = I_{1,a} + I_{1,c}$. 
\vspace{10 mm}

\subsection{Full nonlinear Equations of Motion for the Josephson Parametric Amplifier} \label{subsec:full_order_EOM}
In this subsection, we combine the circuit model for JPA with the input-output relations to construct the full nonlinear EOMs of the JPA. We will take node $1$ as an illustrative example and then give the full set of EOMs for the circuit. Note that the left hand side of the EOM for the closed circuit model of the JRM in Eq.~\eqref{eq:close_node1} is equivalent to the current relation at node $1$, except for a constant factor $\phi_0$. To construct the EOM for the open circuit with all the driving ports, we should take the net current injected into node $1$ to replace the right hand side of the Eq.~\eqref{eq:close_node1}. Applying this procedure to all nodes we obtain the EOMs
\begin{widetext}
\begin{subequations}
\label{eq:node_EOM}
\begin{align}
    \ddot{\varphi_1} & 
        + \frac{\left(3 \varphi_1 - \varphi_2 - \varphi_3- \varphi_4 \right)}{8 C_a L_\textrm{in}}  
        + \frac{1}{2 C_a L_\text{J}}\left[ 
            \sin\left( \varphi_1 - \varphi_2 + \frac{\varphi_\textrm{ext}}{4}\right) - 
            \sin\left( \varphi_4 - \varphi_1 + \frac{\varphi_\textrm{ext}}{4}\right)\right] 
        = \frac{1}{2 C_a \phi_0}\left( I_{1,a} + I_{1,c}\right), \\
    \ddot{\varphi_2} & 
        + \frac{\left(3 \varphi_2 - \varphi_1 - \varphi_3- \varphi_4 \right)}{8 C_b L_\textrm{in}}  
        + \frac{1}{2 C_b L_\text{J}}\left[ 
            \sin\left( \varphi_2 - \varphi_3 + \frac{\varphi_\textrm{ext}}{4}\right) - 
            \sin\left( \varphi_1 - \varphi_2 + \frac{\varphi_\textrm{ext}}{4}\right)\right] 
        = \frac{1}{2 C_b \phi_0}\left( I_{2,b} + I_{2,c}\right), \\
    \ddot{\varphi_3} & 
        + \frac{\left(3 \varphi_3 - \varphi_1 - \varphi_2- \varphi_4 \right)}{8 C_a L_\textrm{in}}  
        + \frac{1}{2 C_a L_\text{J}}\left[ 
            \sin\left( \varphi_3 - \varphi_4 + \frac{\varphi_\textrm{ext}}{4}\right) - 
            \sin\left( \varphi_2 - \varphi_3 + \frac{\varphi_\textrm{ext}}{4}\right)\right] 
        = \frac{1}{2 C_a \phi_0}\left( I_{3,a} + I_{3,c}\right), \\
    \ddot{\varphi_4} & 
        + \frac{\left(3 \varphi_4 - \varphi_1 - \varphi_2- \varphi_3 \right)}{8 C_b L_\textrm{in}}  
        + \frac{1}{2 C_b L_\textrm{J}}\left[ 
            \sin\left( \varphi_4 - \varphi_1 + \frac{\varphi_\textrm{ext}}{4}\right) - 
            \sin\left( \varphi_3 - \varphi_4 + \frac{\varphi_\textrm{ext}}{4}\right)\right] 
        = \frac{1}{2 C_b \phi_0}\left( I_{4,b} + I_{4,c}\right).
\end{align}
\end{subequations}
where the net currents injected from each of the ports to the corresponding nodes are given in Eqs.~\eqref{eq:IO_aI},~\eqref{eq:IO_b} and~\eqref{eq:IO_c}. Using the transformation of Eq.~\eqref{eq:model_matrix} we obtain the EOMs using the normal modes
\begin{subequations}
\label{eq:mode_EOM}
\begin{align}
 \ddot{\varphi}_a + \gamma_a \dot{\varphi}_a + \frac{\varphi_a}{2C_a L_\textrm{in}} + 
     \frac{2}{C_a L_\textrm{J}} & \left[ 
            \sin\left( \frac{\varphi_a}{2} \right) 
            \cos\left( \frac{\varphi_b}{2} \right) 
            \cos\left( \varphi_c \right) 
            \cos\left( \frac{\varphi_\textrm{ext}}{4}\right) \right. \nonumber \\
     - & \left.
            \cos\left( \frac{\varphi_a}{2} \right) 
            \sin\left( \frac{\varphi_b}{2} \right) 
            \sin\left( \varphi_c \right) 
            \sin\left( \frac{\varphi_\textrm{ext}}{4}\right) \right] = 2 \gamma_a \partial_t \varphi_{a,\textrm{in}}(t) \\
 \ddot{\varphi}_b + \gamma_b \dot{\varphi}_b + \frac{\varphi_b}{2C_b L_\textrm{in}} + 
     \frac{2}{C_b L_\textrm{J}} & \left[ 
            \cos\left( \frac{\varphi_a}{2} \right) 
            \sin\left( \frac{\varphi_b}{2} \right) 
            \cos\left( \varphi_c \right) 
            \cos\left( \frac{\varphi_\textrm{ext}}{4}\right) \right. \nonumber \\
     - & \left.
            \sin\left( \frac{\varphi_a}{2} \right) 
            \cos\left( \frac{\varphi_b}{2} \right) 
            \sin\left( \varphi_c \right) 
            \sin\left( \frac{\varphi_\textrm{ext}}{4}\right) \right] = 2 \gamma_b \partial_t \varphi_{b,\textrm{in}}(t) \\
 \ddot{\varphi}_c + \gamma_c \dot{\varphi}_c + \frac{\varphi_c}{C_c L_\textrm{in}} + 
     \frac{4}{C_c L_\textrm{J}} & \left[ 
            \cos\left( \frac{\varphi_a}{2} \right) 
            \cos\left( \frac{\varphi_b}{2} \right) 
            \sin\left( \varphi_c \right) 
            \cos\left( \frac{\varphi_\textrm{ext}}{4}\right) \right. \nonumber \\
     - & \left.
            \sin\left( \frac{\varphi_a}{2} \right) 
            \sin\left( \frac{\varphi_b}{2} \right) 
            \cos\left( \varphi_c \right) 
            \sin\left( \frac{\varphi_\textrm{ext}}{4}\right) \right] = \sqrt{2} \gamma_c \partial_t \varphi_{c,\textrm{in}}(t)
\end{align}
\end{subequations}
\end{widetext}
where we define the effective capacitance for the $c$ mode as $C_c = \frac{4 C_a C_b}{C_a+C_b}$. The mode decay rates $\gamma_a$, $\gamma_b$ and $\gamma_c$ are given by $\gamma_a = \left(C_a Z_a\right)^{-1}$, $\gamma_b = \left(C_b Z_b\right)^{-1}$ and $\gamma_c = \frac{C_a+C_b}{2 C_a C_b Z_c}$. We convert the input-output relations of Eqs.~\eqref{eq:IO_aV},~\eqref{eq:IO_b_V} and~\eqref{eq:IO_c_V} into input-output relations for flux
\begin{subequations}
\label{eq:fluxInOut}
    \begin{align}
        \varphi_{a,\text{out}} & = \varphi_a - \varphi_{a,\textrm{in}} \\ 
        \varphi_{b,\text{out}} & = \varphi_b - \varphi_{b,\textrm{in}} \\
        \varphi_{c,\text{out}} & = \sqrt{2} \varphi_c - \varphi_{c,\textrm{in}}.
    \end{align}
\end{subequations}
The response of the JPA can be fully described using Eqs.~\eqref{eq:mode_EOM} and \eqref{eq:fluxInOut}. 

Finally, we point out that it is useful to use the normal modes of the JRM as the coordinates for writing the EOMs as it makes the analysis of the effects of the various orders of nonlinear coupling easier to understand. On the other hand, using the node fluxes as coordinates is useful as they are more naturally connected to Kirchhoff's law, especially when we want to include experimental imperfections.

\section{Saturation Power of a Josephson Parametric Amplifier (with participation ratio $p=1$)} \label{sec:optimization}

In this section, we first obtain the saturation power of the JPA as described by the exact nonlinear EOMs discussed in subsection~\ref{subsec:full_order_EOM}. Next, we analyze how higher-order nonlinear couplings affect the dynamics of the JPA with the goal of understanding which couplings control the saturation power of the parametric amplifier, to give us guidance on how to improve the saturation power.

We begin with Subsec.~\ref{subsec:full_order}, in which we summarize our main results concerning the dependence of the saturation power on the parameter 
\begin{equation}
    \beta = L_\text{J}/L_\text{in}.
    \label{eq:beta}
\end{equation}
Specifically, we compare numerical solution of the full nonlinear model with numerical solutions of truncated models as well as perturbation theory results. We show that for small $\beta$ the limitation on saturation power comes from dynamically generated Kerr-like terms, while for large $\beta$ saturation power is limited by 5th order non-linearities of the JRM. The details of the analytical calculations are provided in the following subsections.

In Subsec.~\ref{subsec:StP_3rd}, we remind ourselves of the exact analytical solution for the ideal third order amplifier in which the signal is so weak that it does not perturb the pump (i.e. the stiff-pump case). Next, in Subsec.~\ref{subsec:SoP_3rd} we consider the case of a third order amplifier with input signal sufficiently strong such that it can affect the pump (i.e. the soft pump case). In this Subsection we construct a classical perturbation expansion (in which the stiff pump solution corresponds to the zeroth order solution and the first order correction) and find that it leads to the generation of an effective cross-Kerr term, and a pair of two-photon loss terms, one of which could be thought of as and imaginary cross-Kerr term. In Subsec.~\ref{subsec:StP_4th}) we compare the effects of the dynamically generated terms to intrinsic Kerr terms. We analyze fifth and higher order couplings in Subsec.~\ref{subsec:higher_order}).

A note about notation: Throughout this section, we refer to the $a$-mode as the signal mode, $b$-mode as the idler mode, and $c$-mode as the pump mode with intrinsic frequencies $\omega_a$, $\omega_b$ and $\omega_c$. To simplify the discussion of the perturbative expansion, we only consider the case in which we assume that (1) the parametric amplifier is on resonance, i.e. $\delta = \omega_S-\omega_a = 0$ (where $\omega_S$ is the frequency of the signal tone) and $\varepsilon_p = \omega_P - (\omega_a+\omega_b) = 0$ (where $\omega_P$ is the pump tone frequency), so that $\omega_S = \omega_a$, $\omega_I = \omega_b$ and $\omega_P = \omega_a + \omega_b$, (2) the magnetic flux bias is set to the Kerr nulling point, i.e. $\varphi_\textrm{ext} = 2\pi$, (3) an input tone is only sent to the signal mode and there is no input to the idler mode.

\subsection{Main result: saturation power as a function of $\beta$} 
\label{subsec:full_order}
In this subsection, we will compare the exact numerical solution of the full nonlinear EOMs of the JPA to various approximate solutions in order to identify the effects that limit saturation power.

For concreteness, we fix the following parameters: The magnetic field bias is fixed at $\varphi_{\textrm{ext}} = 2\pi$, the mode frequencies are fixed at $\omega_a / (2\pi) = 7.5$~GHz and $\omega_{b}/(2\pi) = 5.0$~GHz ($\omega_c/(2\pi)=6.37$~GHz is fixed by the JPA circuit), the decay rates of the modes are fixed at $\gamma_a/(2\pi) = \gamma_b/(2\pi) = \gamma_c/(2\pi) =  0.1$~GHz, and the critical current of the Josephson junctions is fixed at $i_c=1\,\mu \text{A}$. Throughout, we will set the amplitude of the pump to achieve 20dB reflection gain (at small signal powers). This leaves us with one independent parameter: the JRM inductance ratio $\beta$. 

We shall now analyze saturation of the amplifier as a function of $\beta$. We find it convenient to use saturation input signal flux $\left\vert \varphi_{a,\textrm{in}}(\omega_a) \right\vert$ as opposed to $P_{\pm 1 \textrm{dB}}$ because the former saturates to a constant value at high $\beta$ while the latter grows linearly at high $\beta$. We note that the two quantities are related by the formula
\begin{equation}
        P_{\pm 1 \textrm{dB}} = \frac{\phi_0^2}{2 Z_a} \left\vert\partial_t^2 \varphi_{a,\textrm{in}}(t)\right\vert^2 = \frac{1}{2} C_a \phi_0^2 \gamma_a \omega_a^2 \left\vert \varphi_{a,\textrm{in}}(\omega_a) \right\vert^2.
        \label{eq:sat_power}
\end{equation}

At the nulling point, the EOMs do not explicitly depend on the Josephson junction critical current $i_c$. Therefore, the dynamics of the circuit in terms of the dimensionless fluxes $\varphi_a$, $\varphi_b$, and $\varphi_c$ are invariant if we fix $\omega_a$, $\omega_b$, $\gamma_a$, $\gamma_b$, $\gamma_c$ and $\beta$. However, $i_c$ is needed to connect the dimensionless fluxes to dimensional variables. Specifically, the connection requires the mode capacitance, see Eq.~\eqref{eq:sat_power}. At the nulling point, the mode capacitance is set by the mode frequency and $L_\textrm{in}$ [e.g. $ C_a = 1/(\omega_a^{2} L_\textrm{in})$] and $L_\textrm{in}$ is set by $i_c$, see Eq~\eqref{eq:beta}. In the following, we will analyze saturation power in terms of dimensionless fluxes.

\begin{figure}
    \centering
    \includegraphics[width = 3.2 in]{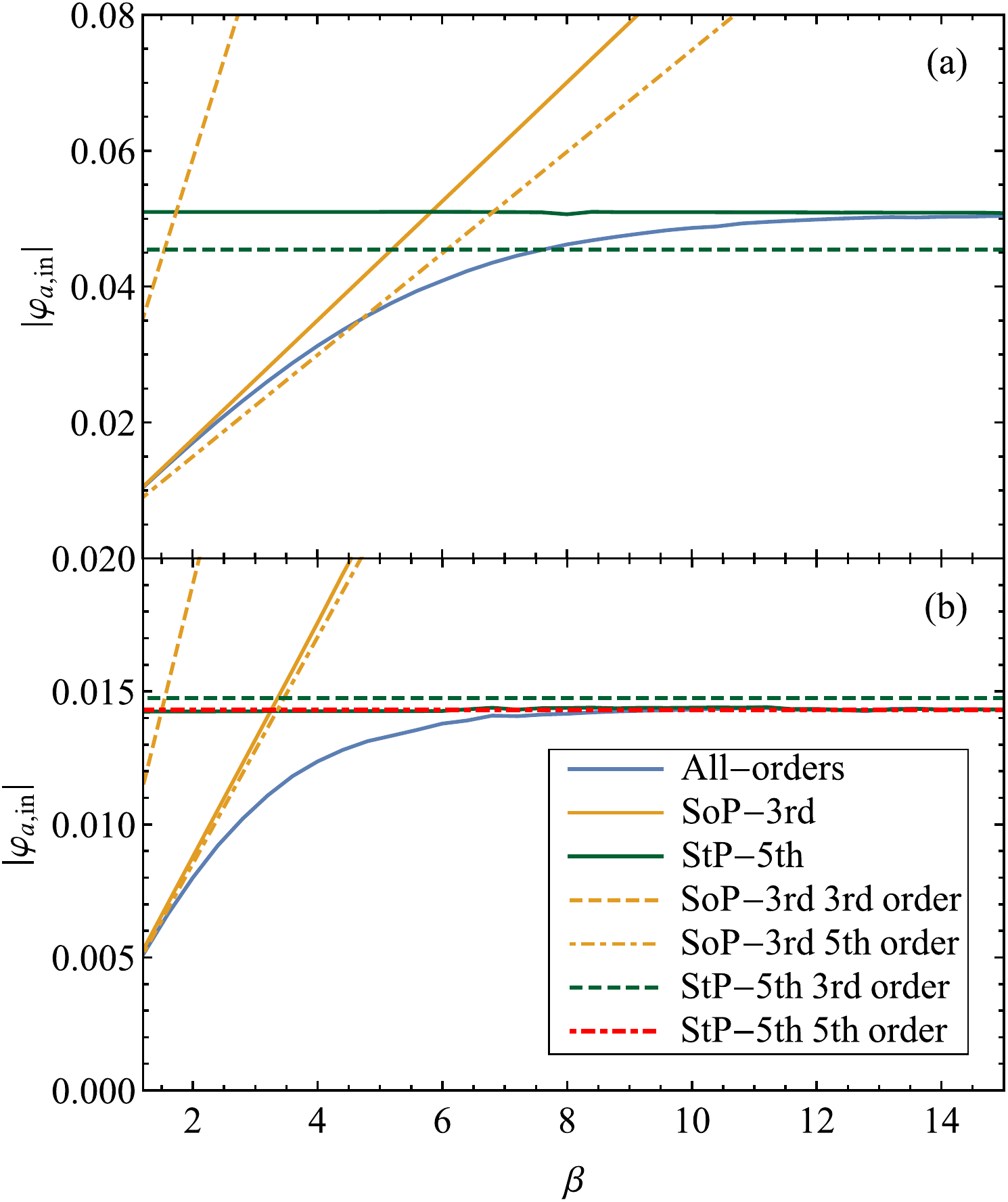}
    \caption{We plot the saturation flux $\vert \varphi_{a,\textrm{in}}\vert $ of the JPA as we change JRM inductance ratio $\beta = L_\textrm{J} / L_\textrm{in}$. The amplifier saturates to $19$~dB and $19.9$~dB in (a) and (b) respectively. The saturation flux from numerical integration of full nonlinear EOMs, SoP-3rd order and StP-5th order nonlinear models of JPA is plotted as blue, orange and dark green solid lines in both subplots. The saturation fluxes are also obtained by the perturbation analysis for the SoP-3rd order and StP-5th order nonlinear models. We plot the third order and fifth order perturbation results as dashed lines and dash-dotted lines. In (a), the perturbation saturation fluxes do not agree well with the numerical ones. This is because the saturation fluxes are already out of the radius of convergence of the perturbation series. While in (b), they have a good agreement. Parameters chosen: $\varphi_{\textrm{ext}} = 2\pi$, the mode frequencies $\omega_a/(2\pi) = 7.5$~GHz, $\omega_b/(2\pi) = 5.0$~GHz, mode decay rates $\gamma/(2\pi) = 100$~MHz for all three modes. The critical current is set to $i_c = 1~\mu$A. We tune the inner shunted inductance $L_{\textrm{in}}$ to tune $\beta$. }
    \label{fig:sweep_beta}
\end{figure}

In Fig~\ref{fig:sweep_beta} we plot the saturation flux as a function of $\beta$ obtained using the full nonlinear EOMs as well as various truncated EOMs and perturbation theory. In panel (a) of the figure we use the conventional criteria that saturation occurs when the gain change by $\pm 1$~dB, while in panel (b) we use the tighter condition that gain changes by $\pm 0.1$~dB. We observe that the saturation flux has two different regimes. At small $\beta$ the saturation flux grows linearly with $\beta$, while at high $\beta$ it saturates to a constant.

To understand the limiting mechanisms in both $\beta$ regimes, we compare the saturation flux $\varphi_{a,\textrm{in}}$ obtained from the numerical integration of full nonlinear EOMs with the various truncated EOMs. We mainly focus on two nonlinear truncated models: (1) soft-pump third order truncated model (SoP-3rd), in which the EOMs of the amplifier are obtained by truncating the Josephson energy to 3rd order in mode fluxes; (2) the stiff-pump fifth order truncated model (StP-5th), in which the Josephson energy is truncated to 5th order in mode fluxes and we ignore the back-action of the signal and idler modes on pump mode dynamics. 

We begin by considering the small $\beta$ regime. Comparing the saturation flux obtained from the numerical integration of full nonlinear EOMs with the above two truncated EOMs, we see that the saturation flux $\varphi_{a,\textrm{in}}$ of the full nonlinear EOMs most closely matches the EOMs of SoP-3rd order model of the amplifier, which indicates the soft-pump condition is the dominating limitation in this regime.

In the soft-pump model, saturation power is limited by the dynamically generated Kerr term. This term shifts the signal and idler modes off resonance as the power in these modes builds up. We describe the details of this process in Subsections~\ref{subsec:SoP_3rd} and \ref{subsec:StP_4th}. In the small $\beta$ regime, the saturation flux of the amplifier increases as we increase $\beta$ [see Fig.~\ref{fig:sweep_beta}(a)]. This is because increasing $\beta$ effectively decreases the nonlinear coupling strength of the amplifier and therefore decreasing the effective strength of the dynamically generated Kerr term. This conclusion is supported by comparing (see Fig.~\ref{fig:sweep_beta}) the exact numerics on the SoP-3rd model (labeled SoP-3rd) with a perturbative analysis of the same model which captures the generated Kerr terms (labeled SoP-3rd 5th order).

In large $\beta$ regime, the saturation flux obtained from full nonlinear model saturates to a constant value (see Fig.~\ref{fig:sweep_beta}, ``All-order'' line). This behavior diverges from the prediction of the SoP-3rd order nonlinear model (``SoP-3rd'' line) but it is consistent with the StP-5th order nonlinear model (``StP-5th" line), which indicates that the dominating limitation in the large $\beta$ regime is the intrinsic 5th order nonlinearity of the JRM energy. Perturbation theory analysis of the StP-5th order nonlinear model (``StP-5th 3rd order" and ``StP-5th 5th order" lines, see Subsec.~\ref{subsec:higher_order}) indicates that the saturation flux depends on the ratio of the fifth order and the third order nonlinear couplings arising from the Josephson non-linearity, and is therefore independent of $\beta$. As we increase $\beta$, the limitation on the saturation power placed by the generated cross-Kerr couplings decreases and hence the mechanism limiting the amplifier's saturation flux changes from generated cross-Kerr couplings to fifth order nonlinearity of the JRM energy. The $\beta$ at which the mechanism controlling saturation flux changes is controlled by the decay rates as $\beta \propto \gamma^{-1/2}$. For our choice of parameters this change of mechanism occurs at $\beta \sim 6$.

\subsection{Ideal parametric amplifier, 3rd order coupling with stiff pump approximation}\label{subsec:StP_3rd}

\begin{figure}[tbp]
    \centering
    \includegraphics[width = 3.2 in]{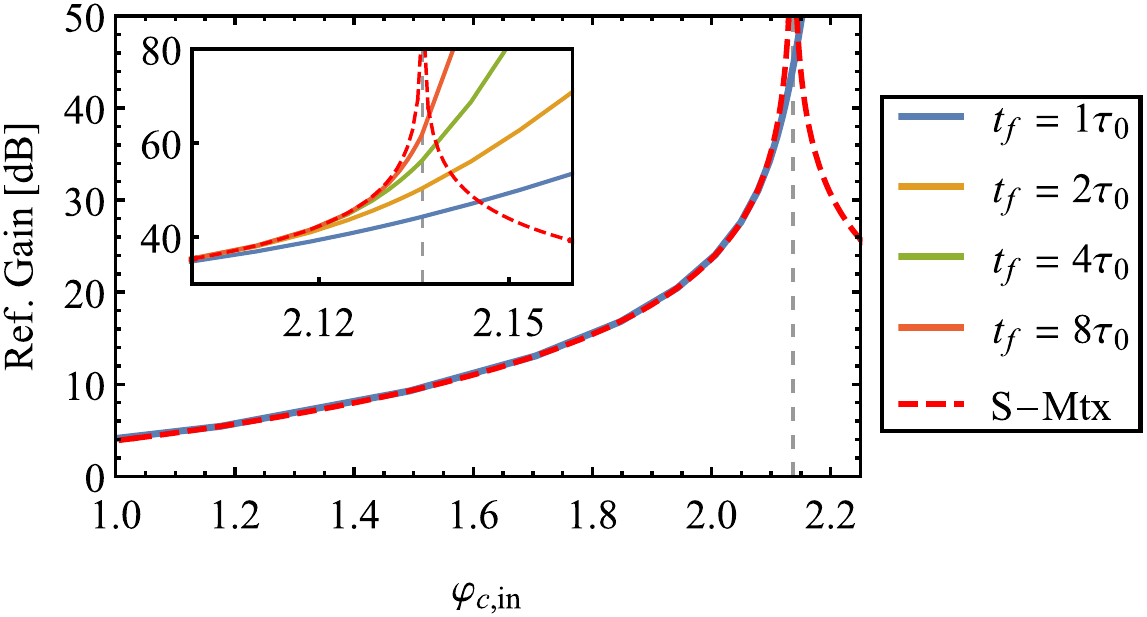}
    \caption{The reflection gain of an ideal parametric amplifier calculated by both scattering matrix (blue curve) and time-domain numerical evolution (red dashed line). The response of the ideal amplifier can be faithfully simulated by the time-domain numerical method with a reasonable reflection gain. The black dashed line shows the bias point beyond which the amplifier is unstable. Our time-domain numerical solution start to deviated form the scattering matrix calculation at $\sim 2.1$. This is because the numerical accuracy of the time-domain solver. In the insert, we increase the final time $t_f$ of the time-solver. We notice a better and better convergence to the analytical solution (red dashed line).
    This is caused by a numerical instability that occurs near the divergence point of the amplifier, Eq.~\eqref{eq:large_gain}. Parameters chosen: $\varphi_\textrm{ext} = 2\pi$, $\omega_a/(2\pi) = 7.5$~GHz, $\omega_b/(2\pi) = 5.0$~GHz, $\gamma_a = \gamma_b = \gamma_c = 2\pi \times 0.01$~GHz and $i_c = 1.0~\mu$A. Time constant $\tau_0 = 4000 f_a^{-1}$ where $f_a$ is the signal mode frequency.}
    \label{fig:ideal_para_amp}
\end{figure}

In this subsection, we remind ourselves with the solution of ideal parametric amplifier. The ideal parametric amplifier can be exactly solved in frequency domain such that we can also verify the reliability of the numerical solutions.

In an ideal parametric amplifier, the only coupling present is a third order coupling between the signal, idler and pump mode that results in parametric amplification. Further, the pump mode strength is considered to be strong compared to the power consumed by the amplification, such that the pump mode dynamics can be treated independently of the signal and idler modes. This approximation is commonly referred to as the ``stiff-pump approximation'' (StP). The EOMs to describe the parametric amplifier can be derived from the full nonlinear EOMs in Eq.~\eqref{eq:mode_EOM} by expanding the nonlinear coupling terms to second order in mode fluxes $\varphi$'s (2nd order in EOMs corresponding to 3rd order in Lagrangian). Under the stiff-pump approximation, we can effectively remove the three mode coupling terms in the EOM for the pump mode
\begin{equation}
 \ddot{\varphi}_c  + \gamma_c \dot{\varphi}_c + \omega_{c}^2 \varphi_{c} = \sqrt{2} \gamma_c \partial_t \varphi_{c,\textrm{in}}(t).
 \label{eq:EOM_StP_3rd_c}
\end{equation}
$\varphi_{c}$ obtained from this equation acts as a time-dependent parameter in the EOMs for the $a$ and $b$ modes
\begin{subequations}
    \label{eq:EOM_StP_3rd}
\begin{align}
 \ddot{\varphi}_a & + \gamma_a \dot{\varphi}_a + \omega_{a}^2 \varphi_{a} - 
     \frac{2\omega_{a}^2}{\beta} \varphi_b \varphi_c
        = 2 \gamma_a \partial_t \varphi_{a,\textrm{in}}(t) \label{eq:EOM_StP_3rd_a},\\
 \ddot{\varphi}_b & + \gamma_b \dot{\varphi}_b + \omega_{b}^2 \varphi_{b} -
     \frac{2\omega_{b}^2}{\beta} \varphi_a \varphi_c
         = 2 \gamma_b \partial_t \varphi_{b,\textrm{in}}(t) \label{eq:EOM_StP_3rd_b}.
\end{align}
\end{subequations}
Assuming the pump tone is $\varphi_{c,\textrm{in}} (t)  = \varphi_{c,\textrm{in}} e^{-i\omega_P t} + c.c.$, we find that $\varphi_c (t) = \varphi_c (\omega_P) e^{-i \omega_P t} + c.c.$ where,
\begin{equation}
    \varphi_c (\omega_P) = \frac{-i \sqrt{2} \gamma_c \omega_P}{\omega_c^2 - \omega_P^2 -i\gamma_c \omega_P} \varphi_{c,\textrm{in}}
    \label{eq:StP_c_mode}
\end{equation}
Substituting the $c$-mode flux $\varphi_c$ in Eq.~\eqref{eq:EOM_StP_3rd} the EOMs for $a$ and $b$ modes become linear and can be solved in the frequency domain. The Fourier components of the $a$ and $b$ modes, under the rotating-wave approximation, are
\begin{subequations}
\label{eq:stpSignalIdler}
    \begin{align}
        \varphi_a (\omega_a) = \frac{2 \tilde{\gamma}_a \tilde{\gamma}_b}{\tilde{\gamma}_a \tilde{\gamma}_b - 4 g^{2} \left\vert \varphi_c (\omega_P) \right\vert^2} \varphi_{a,\textrm{in}} \\
        \varphi_b^{*} (\omega_b) = \frac{4 i g \tilde{\gamma}_a \varphi_c^{*} (\omega_P) }{\tilde{\gamma}_a \tilde{\gamma}_b - 4 g^{2}\left\vert \varphi_c (\omega_P) \right\vert^2} \varphi_{a,\textrm{in}}
    \end{align}
\end{subequations}
where we define the dimensionless decay rates $\tilde{\gamma}_j = \gamma_j / \omega_j$, the dimensionless three-mode coupling strength $g = (1/\beta) \sin \left( \frac{\varphi_\textrm{ext}}{4}\right) = 1/\beta$, which is obtained from a series expansion of the dimensionless potential energy 
\begin{align}
    \mathcal{E_\textrm{JRM}} \equiv \left[ (\phi_0^2 / L_{\textrm{in}})\right]^{-1} E_{\textrm{JRM}}.
    \label{eq:Edimless}
\end{align}
Here we assume the input tone is $\varphi_{a,\textrm{in}} = \varphi_{a,\textrm{in}} e^{-i \omega_a t} + c.c.$ and there is no input into idler ($b$) mode. 

The linear response of the ideal parametric amplifier is obtained using scattering matrix formalism. The EOMs of an ideal parametric amplifier can be written in matrix form as $[M].[\varphi] = 2 [\tilde{\gamma}].[\varphi_{\textrm{in}}]$, where 
\begin{align}
    [M] &= \left( 
        \begin{array}{cc}
            \tilde{\gamma}_a  &  -2 i g \varphi_c(\omega_P) \\
            2 i g \varphi_c^{*}(\omega_P) &  \tilde{\gamma}_b
        \end{array}
    \right), \label{eq:M} \\
    [\tilde{\gamma}] &= \left( 
        \begin{array}{cc}
            \tilde{\gamma}_a & \\
             & \tilde{\gamma}_b
        \end{array}
    \right).
    \label{eq:on_resonance_mtx}
\end{align}

The scattering matrix, which is defined by $[\varphi_{\textrm{out}}] = [S].[\varphi_{\textrm{in}}]$, is given by $[S] = 2  [M^{-1}].[\tilde{\gamma}] - I_{2\times2}$, where we have used the input-output relation $[\varphi]=[\varphi_{\text{in}}]+[\varphi_{\text{out}}]$ and $I_{2\times2}$ is the $2\times2$ identity matrix,
\begin{equation}
    [S] = 
        \left(
            \begin{array}{cc}
                \frac{2 \tilde{\gamma}_a \tilde{\gamma}_b}{\tilde{\gamma}_a \tilde{\gamma}_b - 4 g^{2} \left\vert \varphi_c (\omega_P) \right\vert^2} -1 &  - \frac{4 i g \tilde{\gamma}_b \varphi_c (\omega_P)}{\tilde{\gamma}_a \tilde{\gamma}_b - 4 g^{2}\left\vert \varphi_c (\omega_P) \right\vert^2} \\
                \frac{4 i g \tilde{\gamma}_a \varphi_c^{*} (\omega_P) }{\tilde{\gamma}_a \tilde{\gamma}_b - 4 g^{2}\left\vert \varphi_c (\omega_P) \right\vert^2} & \frac{2 \tilde{\gamma}_a \tilde{\gamma}_b}{\tilde{\gamma}_a \tilde{\gamma}_b - 4 g^{2} \left\vert \varphi_c (\omega_P) \right\vert^2} -1 
            \end{array}
        \right).
        \label{eq:sMtx_simp}
\end{equation}
The reflection gain of the signal mode (in units of power) is defined as $G_0 = \left\vert [S]_{11} \right\vert^2$. To get large gain ($G_0 \gg 1$), the pump mode strength should be tuned to 
\begin{equation}
    2 g \left\vert \varphi_c (\omega_P) \right\vert \sim \sqrt{\tilde{\gamma}_a \tilde{\gamma}_b}.
    \label{eq:large_gain}
\end{equation}

Alternatively, we can obtain the response of the JPA using time-domain numerical integration. First, we solve for the mode variables inside the JPA circuit with specific signal and pump inputs. Next, we use the input-output relation to find the output signal and then we obtain the reflection gain of the amplifier. Specifically, to solve the dynamics of the parametric amplifier, we set the input signal as $\varphi_{a,\text{in}}(t) = \bar{\varphi}_{a,\text{in}} \cos (\omega_S t)$ and $\varphi_{b,\text{in}} = 0$, and numerically integrate the EOMs [Eq.~\eqref{eq:EOM_StP_3rd}]. Here we note that $\bar{\varphi}_{a,\text{in}} = 2 \varphi_{a,\text{in}}$, which is defined in Eq.~\eqref{eq:stpSignalIdler}. In Fig.~\ref{fig:ideal_para_amp}, we show the comparison of the reflection gain obtained using numerical integration (red dashed line) and the scattering matrix solution (blue solid line). The two solutions start out identical. However, as we increase the pump mode strength $\varphi_{c,\text{in}}$, we notice that as the reflection gain starts diverging ($G_0\sim 35$~dB, see the insert of Fig.~\ref{fig:ideal_para_amp}) from the analytical solution. This is because the numerical solver need a longer and longer time window to establish the steady-state solution of the nonlinear EOMs as we move towards the unstable point (vertical dashed line). To optimize the run time, here and later in the paper, we choose the time-window for our solver so that the numerical solution saturates for amplification of $\sim20$~dB.

In the unstable regime the reflection flux on the signal mode diverges exponentially with time, as the amplifier will never run out of power under the StP approximation. Therefore,  in this regime the time-domain solver gives a large unphysical reflection gain (as we cut it off at some large, but finite time). 

\begin{figure}[htbp!]
    \centering
    \includegraphics[width = 3.0 in]{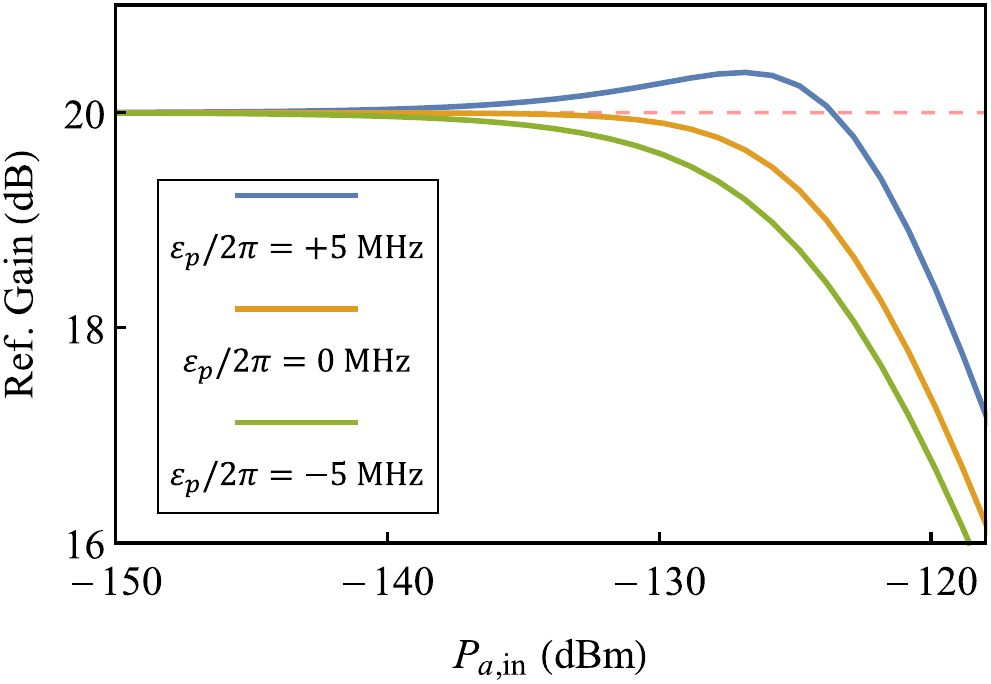}
    \caption{We consider the soft-pump condition with third order coupling strength and calculate the reflection gain of the amplifier. We slightly detune the pump drive frequency from the sum frequency of the signal and idler mode frequency. When the pump frequency detuning $\varepsilon_p$ is negative (green line), the reflection gain is further suppressed compared with on-resonance drive (orange line). However, when the pump frequency detuning is positive (blue line), the ``shark fin'' feature reappears, which was understood as the consequence of the existence of Kerr nonlinearity in the amplifier system.}
    \label{fig:EOM_expansion}
\end{figure}

\subsection{JPA with third order coupling, relaxing the stiff-pump approximation} \label{subsec:SoP_3rd}
As we increase the input signal strength, the power supplied to the pump mode will eventually be comparable to the power consumed by amplification, where the amplifier will significantly deviate from the ideal parametric amplifier. In this subsection, we reinstate the action of the signal and idler modes on the pump mode. Since the pump mode strength is affected by the signal and idler mode strengths, we refer to it as the ``Soft-pump'' (SoP) condition. 

The EOMs for the Soft-pump third order model of the JPA can be obtained by expanding the full nonlinear EOMs [Eq.~\eqref{eq:mode_EOM}] and truncating all three EOMs to second order in mode fluxes. That is, we use Eq.~\eqref{eq:EOM_StP_3rd} to describe $a$ and $b$ modes and modify Eq.~\eqref{eq:EOM_StP_3rd_c} for the $c$ mode as 
\begin{align}
\ddot{\varphi}_c  + \gamma_c \dot{\varphi}_c + \omega_{c}^2 \varphi_{c} - \frac{\omega_c^2}{\beta} \varphi_a \varphi_b = \sqrt{2} \gamma_c \partial_t \varphi_{c,\textrm{in}}(t).
\end{align}
Unlike the StP approximation, the $c$ mode flux $\varphi_c$ can no longer be treated as a time-dependent parameter unaffected by $a$ and $b$ modes. While we can no longer obtain an exact analytical solution to these EOMs, we use perturbation theory as well as time-domain numerical integration to seek the dynamics of the amplifier.

In Fig.~\ref{fig:EOM_expansion}, we plot the reflection gain obtained by numerical integration. The reflection gain of the amplifier is no longer independent of the input signal power, instead we see that the reflection gain deviates from $20$~dB as we increase the signal mode power. Moreover, as we change the detuning of the pump mode relative to the sum frequency of the signal and idler mode, the deviation of the reflection gain changes from negative to positive. While a deviation towards smaller gain (which occurs at negative or zero detuning) is consistent with the pump saturation scenario, a deviation towards higher gain  (which occurs at positive detuning) is not. The ``shark fin'' feature we observe here, in which the gain first deviates up and then down, has been previously attributed to intrinsic Kerr couplings~\cite{GQLiu2017}. The fact that the ``shark fin'' reappears without an intrinsic Kerr term gives us a hint that SoP-3rd order couplings can generate an effective Kerr nonlinearity.

To fully understand the effect of the SoP condition, we use classical perturbation theory to analyze the dynamics of the circuit. Below, we explain the essential steps of the perturbation analysis. Then, we focus on the SoP-3rd order truncated model and compute the parametric dependence of the saturation flux of the amplifier.

\subsubsection{Classical perturbation theory for the Josephson Parametric Amplifier}
The small parameter in our perturbative expansion is the input fluxes to the signal and idler modes, $\varphi_{a,\textrm{in}}$ and $\varphi_{b,\textrm{in}}$. We can expand the mode fluxes in a series as
\begin{equation}
    \varphi_{j}(t) = \varphi_{j}^{(0)}(t) + \varphi_{j}^{(1)}(t) + \varphi_{j}^{(2)}(t) ...\,
\end{equation}
for $j=a, b, c$. The EOM of the signal mode flux $\varphi_a$ after series expansion is
\begin{widetext}
\begin{equation}
     \left( \partial_t^2 + \gamma_a \partial_t + \omega_{a}^2 \right) \left(\varphi_a^{(1)}(t) + \varphi_a^{(2)}(t) + ... \right) - 
         \frac{2\omega_{a}^2}{\beta}
            \left( \varphi_b^{(1)}(t) + \varphi_b^{(2)}(t)+ ... \right)
            \left( \varphi_c^{(0)}(t) + \varphi_c^{(1)}(t)+ ... \right) 
            = 2 \gamma_a \partial_t \varphi_{a,\textrm{in}}(t).
    \label{eq:SoP_3rd_expansion}
\end{equation}
\end{widetext}
The idler and pump mode EOMs are similar. 

In the absence of inputs to the signal and idler modes, we obtain the zeroth order solution of the EOMs. Since the amplifier should be stable, there should be no output in the signal and idler modes when there is no input, i.e. $\varphi_a^{(0)} = \varphi_b^{(0)} = 0$. Therefore, the only nonzero zeroth order solution is for the pump mode, which is given by
\begin{equation}
    \ddot{\varphi}_c^{(0)} + \gamma_c \dot{\varphi}_c^{(0)} + \omega_{c}^2 \varphi_{c}^{(0)} = \sqrt{2} \gamma_c \partial_t \varphi_{c,\textrm{in}}(t).
\end{equation}
This equation matches the StP $c$ mode EOM [see Eq.~\eqref{eq:EOM_StP_3rd_c}], the zeroth-order solution for $\varphi_c$ is given in Eq.~\eqref{eq:StP_c_mode}. We can then solve the higher corrections to signal, idler and pump mode fluxes by matching the terms in the EOMs order-by-order. For example, the equations for first order corrections $\varphi_a^{(1)}$ and $\varphi_b^{(1)}$ are identical to the ideal parametric amplifier, and hence they are given by the StP solution Eq.~\eqref{eq:stpSignalIdler}, while the first order correction to the pump mode flux is $\varphi_c^{(1)}=0$.

As the first order correction to the pump mode is zero, there are no second order corrections to the signal and idler mode fluxes. While the second order correction to the pump mode has two frequency components: $\Sigma = \omega_S+ \omega_I$ and $\Delta = \omega_S- \omega_I$ with Fourier components
\begin{subequations}
    \label{eq:SoP_3rd_phic_2nd}
    \begin{align}
        \varphi_c^{(2)}(\Sigma) & = f_\Sigma \frac{1}{\beta} \varphi_a^{(1)}(\omega_S)\varphi_b^{(1)}(\omega_I) \\
        \varphi_c^{(2)}(\Delta) & = f_\Delta\frac{1}{\beta} \varphi_a^{(1)}(\omega_S)\varphi_b^{(1)*}(\omega_I),
    \end{align}
\end{subequations}
where the two dimensionless parameters $f_{\Sigma}$ and $f_{\Delta}$ are defined as
\begin{subequations}
        \label{eq:f_parameters}
        \begin{align}
           f_{\Sigma}  & = \frac{\omega_{c}^2}{\omega_c^2 - \Sigma^2 -i \gamma_c \Sigma}, \\
           f_{\Delta}  & = \frac{\omega_{c}^2}{\omega_c^2 - \Delta^2 -i \gamma_c \Delta}.
        \end{align}
\end{subequations}
Both of these two frequency components contribute to the third order correction to the signal and idler mode flux with frequency $\omega_S$ and $\omega_I$. 

To obtain the third order corrections to the signal and idler mode fluxes we define an effective drive vector that is comprised of all the contributions from lower orders, utilizing Eq.~\eqref{eq:SoP_3rd_phic_2nd} to express $\varphi_c^{(2)}$ in terms of $\varphi_a^{(1)}$ and $\varphi_b^{(1)}$
\begin{equation}
[\varphi_d^{(3)}] =  
    \left(
     \begin{array}{c}
        2 g^2 \left( f_\Delta + f_\Sigma \right) \varphi_a^{(1)} \left\vert \varphi_b^{(1)} \right\vert^2
         \\
        2 g^2 \left( f_\Delta + f_\Sigma ^{*} \right)
        \varphi_b^{(1)*} \left\vert \varphi_a^{(1)} \right\vert^2
    \end{array}
    \right).
    \label{eq:SoP_3rd_drive}
\end{equation}

The third order correction to the signal and idler mode is given by $[\varphi^{(3)}] = [M^{-1}].[i].[\varphi_d^{(3)}]$, where $[M]$ is the same matrix as in the discussion of the ideal parametric amplifier Eq.~\eqref{eq:M}, and $[i] = \textrm{diag}\{i,-i\}$ is a diagonal $2\times2$ matrix. The signal mode 3rd order correction is
\begin{align}
        \varphi_a^{(3)} 
           = \left(\frac{1}{\beta}\right)^2
           & \left\lbrace 
           2 i [M^{-1}]_{11} \left( f_\Sigma + f_\Delta\right) \varphi_a^{(1)} \left\vert \varphi_b^{(1)}\right\vert^2 \right. \nonumber \\
            -
           & \left. 2 i [M^{-1}]_{21} \left( f_\Sigma^{*} + f_\Delta\right) \varphi_b^{(1)*} \left\vert \varphi_a^{(1)}\right\vert^2
            \right\rbrace.
            \label{eq:phi3}
\end{align}
Using this expression we obtain the corrections to the reflection gain up to second order $G^{(2)} = \vert \varphi_a^{(1)}(\omega_S) + \varphi_a^{(3)}(\omega_S) - \varphi_{a,in}\vert^2 / \vert {\varphi_{a,in}}\vert^2$. Similarly, we can solve the perturbation theory order-by-order till the desired order.

Here we want to stress that we only focus on the main frequency components of signal and idler modes, i.e. $\varphi_a (\omega_a)$ and $\varphi_b (\omega_b)$ and ignore the higher order harmonics. This assumption is also applied when we consider the higher than 3rd order nonlinear couplings in the JPA truncated EOMs, e.g. in StP-Kerr nonlinear truncated model (discussed in subsec.~\ref{subsec:StP_4th}) and StP-5th order truncated model (discussed in subsec.~\ref{subsec:higher_order}). 

Further, we point out that the above discussion is easily generalized to the case when $\omega_S \neq \omega_a$, $\omega_I \neq \omega_b$ and (or) $\varphi_{\textrm{ext}} \neq 2\pi$.

Next, we consider the question, how the perturbation on the reflection gain can be used to compute the saturation power of the amplifier. The saturation power is defined as the input power at which the amplifier's reflection gain changes by $1$~dB. At the limit $\varphi_{a,\textrm{in}} \rightarrow 0$, the reflection gain of the amplifier is noted as $G_0$, which is given by $G_0 = \vert \varphi_a^{(1)} - \varphi_{a,\textrm{in}} \vert^2 / \vert \varphi_{a,\textrm{in}} \vert^2$.

As we increase the input signal strength $\varphi_{a,\textrm{in}}$ to reach $1$~dB suppression of the reflection gain, the corrected gain (in power unit) should satisfy,
\begin{equation}
    G = \frac{
    \left\vert
        \varphi_{a}^{(1)} + \varphi_{a}^{(c)} - \varphi_{a,\textrm{in}}
    \right\vert^2}{
    \left\vert
        \varphi_{a,\textrm{in}}
    \right\vert^2
    } = 10^{-0.1} G_0
    \label{eq:solve_G}
\end{equation}
where $\varphi_a^{(c)}$ is the higher order corrections to the signal mode flux in perturbation theory. In the high gain limit ($G_0 \gg 1$), we can estimate the criteria by,
\begin{equation}
    \left\vert \varphi_{a}^{(c)} \right\vert / \left\vert \varphi_{a}^{(1)}\right\vert = \epsilon \equiv 10^{-0.05} - 1.
    \label{eq:ep}
\end{equation}
Note $\epsilon$ depends on the definition of the threshold for the gain change at the amplifier saturation.

\subsubsection{Perturbative analysis on SoP third order EOM.}
We apply the above perturbation analysis to SoP-3rd order truncated model to understand the mechanism of amplifier saturation in this model. 
Before we proceed to calculate the corrections to the reflection gain, we estimate the matrix elements in the inverse of the parametric matrix $[M]$ [see Eq.~\eqref{eq:M}] in high gain limit, i.e. 
\begin{equation}
\begin{aligned}
        G_A \equiv \sqrt{G_0} & = 2 \tilde{\gamma}_a [M^{-1}]_{11} - 1  \\
        &= \frac{2 \tilde{\gamma}_a \tilde{\gamma}_b}{\tilde{\gamma}_a \tilde{\gamma}_b - 4 g^{2} \left\vert \varphi_c (\omega_P) \right\vert^2} -1 \gg 1.
\end{aligned}
\end{equation}
Therefore, we can approximate $2 \tilde{\gamma}_a [M^{-1}]_{11} \sim G_A$ and hence $\varphi_a^{(1)}(\omega_a) \sim G_A \varphi_{a,\text{in}}$. The matrix element $[M^{-1}]_{21}$ can be approximated by $-i G_A/(2 \sqrt{\tilde{\gamma}_a \tilde{\gamma}_b})$, which can be seen from the relation $[M^{-1}]_{21} =-i 2 g \varphi_c^{(0)*} ([M^{-1}]_{11})/\tilde{\gamma}_b$ and $\sqrt{\tilde{\gamma}_a \tilde{\gamma}_b} \sim 2 g \vert \varphi_c^{(0)} \vert$.

The third order correction to signal mode strength is given by the Eq.~\eqref{eq:phi3}, which becomes 
\begin{equation}
    \varphi_a^{(3)}(\omega_b) 
    \sim 2 \frac{g^2}{\tilde{\gamma}_a} G_A^4 \textrm{Im}(f_\Sigma) \varphi_{a,\text{in}}^3
\end{equation}
in the high gain approximation.

To calculate the saturation flux, we let $\varphi_a^{(3)} \sim \epsilon \varphi_a^{(1)}$ and solve for $\varphi_{a,\textrm{in}}$, where $\epsilon$ is given in Eq.~\eqref{eq:ep}. The saturation flux given by 3rd order perturbation is
\begin{equation}
\varphi_{a,\textrm{in},\pm 1~\textrm{dB}} \sim G_0^{-3/4} \sqrt{\epsilon} \frac{\sqrt{\tilde{\gamma}_b}} {g} \textrm{Im}(f_\Sigma)^{-1/2}.
\label{eq:saturation_flux_SoP_3rd_3rd}
\end{equation}
where $G_0$ is the small-signal reflection gain of the amplifier. The saturation flux given by third order perturbation theory of SoP-3rd nonlinear model is plotted as orange dashed line in Fig.~\ref{fig:sweep_beta}(a)~\footnote{To be more accurate, we directly solve $\varphi_{a}^{(3)} = \epsilon \varphi_{a}^{(1)}$ without high-gain assumption for the perturbation curves in Fig.~\ref{fig:sweep_beta}.}. We notice that the saturation flux predicted by 3rd order perturbation theory does not agree well with the numerical simulation (orange solid line). The disagreement also occurs when we tighten the criteria for amplifier saturation to $0.1$~dB (see Fig.~\ref{fig:sweep_beta}(b) orange dashed line versus orange solid line). 

To explain the disagreement between the perturbation theory and the numerical integration method, we correct the signal mode flux to the next non-zero order, which is at fifth order in $\varphi_{a,\textrm{in}}$. To solve the fifth order correction of signal and idler mode fluxes, we follow the same strategy as demonstrated above. The only nonzero fourth order correction is $\varphi_{c}^{(4)}$, with two frequency components, $\varphi_c^{(4)} (\Sigma)$ and $\varphi_c^{(4)} (\Delta)$. The fifth order correction to the signal mode strength $\varphi_{a}^{(5)}$ is
\begin{equation}
    \varphi_{a}^{(5)} \sim \frac{g^4}{\tilde{\gamma}^2} G_0^{5/2} \textrm{Re}\left[f_\Sigma + f_\Delta\right]^2 \varphi_{a,\textrm{in}}^5
\end{equation}
where we use the fact that imaginary parts of $f_\Delta$ and $f_\Sigma$ are much smaller than their real parts and hance we ignore the contribution from their imaginary parts. The saturation flux can be estimated by $\vert \varphi_a^{(5)} \vert \sim \epsilon \vert \varphi_a^{(1)} \vert$ as, 
\begin{equation}
\varphi_{a,\textrm{in},\pm 1~\textrm{dB}} \sim G_0^{-5/8} \frac{\sqrt{\tilde{\gamma}_b}} {g} \left[ \frac{\epsilon}{2} \textrm{Re}(f_\Sigma + f_\Delta)^{-1} \right]^{1/4}
\label{eq:saturation_flux_SoP_3rd_5th}
\end{equation}

Compared with the saturation flux given by third order perturbation, the fifth order correction is more significant as $f_\Delta$ and $f_\Sigma$ are almost real. However, in third order perturbation theory, the contribution of real parts of $f_\Delta$ and $f_\Sigma$ is canceled, but they will appear in next order perturbation, which dominates the saturation.

The saturation flux correction till fifth order perturbation is obtained by directively solving Eq.~\eqref{eq:solve_G} for $\varphi_{a,\textrm{in}}$, where the corrections of signal mode strength $\varphi_a^{(c)} = \varphi_a^{(3)} + \varphi_a^{(5)}$. The saturation flux corrected upto fifth order [orange dash-dotted line in Fig.~\ref{fig:sweep_beta}(a) and (b)] have better agreement with the numerical solution.

However, in both third order and fifth order perturbation analysis, the saturation flux with $1$~dB gain change does not agree well with the numerical solution [see Fig.~\ref{fig:sweep_beta}(a)]. This is because the saturation flux for $\pm 1$~dB is beyond the radius of convergence of the perturbation series. In order to validate the perturbation analysis, we tight the criteria for amplifier saturation to change of the amplifier gain by$\pm 0.1$~dB, which makes the signal flux to stay in the radius of convergence. In Fig.~\ref{fig:sweep_beta}(b), the saturation flux corrected to fifth order (orange dot-dashed line) has a much better agreement with the numerical methods [orange solid line in Fig.~\ref{fig:sweep_beta}(b)]. 

We notice that the saturation flux is inversely proportional to $g = 1/\beta$, and hence we expect that it can be increased by decreasing the three-mode coupling strength (increasing $\beta$). At the same time, the pump strength must be increased in order to reach $G_0$. This procedure, in effect, makes the pump stiffer. 

\begin{figure*}[htbp!]
    \centering
    \includegraphics[width = 6.0 in]{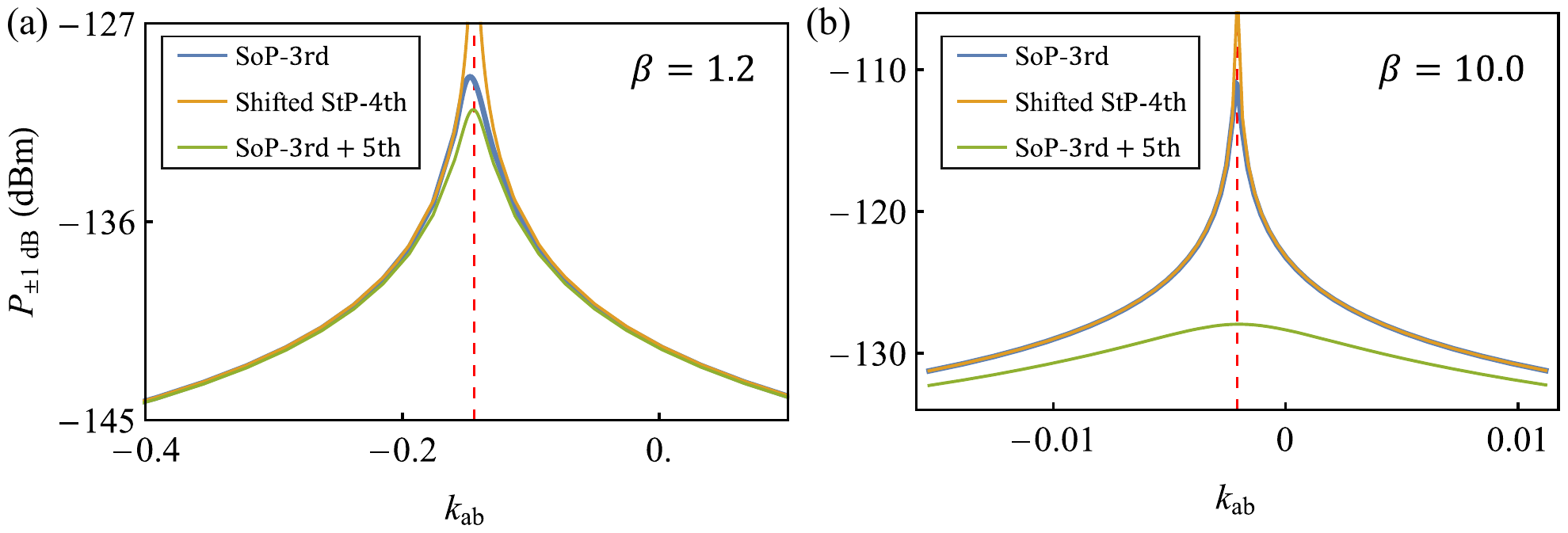}
    \caption{We compare three different cases, soft-pump with third order coupling (SoP-3rd), Stiff-pump with Kerr coupling (StP-4th) and soft pump truncated till the third order with fifth order couplings between signal and idler mode (SoP-3rd + 5th). For each cases, we manually turn on a cross-Kerr coupling $k_{ab}$. For StP-4th case, the plot is shifted by $-k_{ab}^{(\textrm{eff})}$. The vertical red dashed line shows the location where the real part of the dynamically generated cross-Kerr is fully compensated by the intrinsic cross-Kerr coupling. The parameters chosen:  $\omega_a/(2\pi) = 7.5$~GHz, $\omega_b/(2\pi) = 5.0$~GHz, $\gamma_j/(2\pi) = 100$~MHz. We set $\beta = 1.2$ [in (a)] and $10.0$ [in (b)]. The critical current is $i_c = 1.0~\mu$A.
    }
    \label{fig:Kerr_compensation}
\end{figure*}

\subsection{Intrinsic and Generated Kerr Couplings} \label{subsec:StP_4th}
In this subsection, we comment on the generation of effective Kerr terms and compare it with the intrinsic cross-Kerr couplings in the Lagrangian. In the perturbation analysis, if we expand the pump mode strength to second order, the effective EOMs of the signal and idler modes contains a cross Kerr coupling term, in the form of $\varphi_a |\varphi_b|^2$ for the signal mode and $|\varphi_a|^2 \varphi_b$ for the idler mode (see, e.g. Eq.~\eqref{eq:phi3}). We will show that these generated Kerr terms limit the saturation power (at least for small $\beta$). 

To construct an understanding of this mechanism, we use perturbation theory to analyze the StP amplifier with an intrinsic cross Kerr $k_{ab}$, and compare it with the SoP-3rd order nonlinear amplifier. As we discussed in subsection~\ref{subsec:StP_3rd}, in the stiff pump approximation, we treat the pump mode flux, $\varphi_c$, as a time-dependent parameter that is independent of the signal and idler modes. The EOMs for the signal and idler modes can be obtained by adding the terms $4 k_{ab} \varphi_a \varphi_b^2$ and $4 k_{ab} \varphi_a^2 \varphi_b$ to the left-hand-side of Eq.~\eqref{eq:EOM_StP_3rd_a} and~\eqref{eq:EOM_StP_3rd_b}, respectively. 

In perturbation analysis, following the discussion in the previous subsection, we expand the signal and idler mode fluxes in the order of $\varphi_{a,\textrm{in}}$ and $\varphi_{b,\textrm{in}}$. We further assume that the amplifier is stable, i.e. there is no output from the amplifier if there is no input, which gives the zeroth order solution of signal and idler modes as $\varphi_a^{(0)} = \varphi_b^{(0)} = 0$. The first order solution of signal and idler mode fluxes repeats the solution of ideal parametric amplifier [Eq.~\eqref{eq:stpSignalIdler}] and the next non-zero correction appears at third order. The corresponding drive term is
\begin{equation}
    \begin{aligned}
        [\varphi_d^{(3)}] = \left( 
            \begin{array}{c}
            8 k_{ab} \varphi_a^{(1)}(\omega_S) \left\vert \varphi_b^{(1)}(\omega_I) \right\vert^2 \\
            8 k_{ab} \varphi_b^{(1)*} (\omega_I) \left\vert \varphi_a^{(1)}(\omega_S) \right\vert^2
            \end{array}
        \right).
    \end{aligned}
\end{equation}
Comparing with Eq.~\eqref{eq:SoP_3rd_drive}, we see that the soft-pump condition gives an effective signal-idler Kerr coupling strength
\begin{equation}
    k_{ab}^{\textrm{eff}} = \frac{1}{4} g^2 \left[ f_\Delta + \textrm{Re}(f_\Sigma) \right].
    \label{eq:keff}
\end{equation}
We note that this effective Kerr coupling is complex as $f_\Delta$ is complex. We also observe that there is an additional term in Eq.~\eqref{eq:SoP_3rd_drive}, that we label $q_{ab}^{\text{eff}}= (1/4) g^2 \text{Im} (f_\Sigma)$, which cannot be mapped onto a Kerr coupling (as the signal and idler parts have opposite sign).

Further, as the intrinsic cross Kerr coupling $k_{ab}$ is real, the third order correction to the signal mode in StP cross-Kerr amplifier model is zero. If we proceed to next non-zero order correction to the signal and idler mode, and compare the drive term with the one from StP-3rd order truncated model in same perturbation order, we identify the same effective Kerr coupling strength as Eq.~\eqref{eq:keff}. 

To check the correspondence and understand to what degree the saturation power of SoP-3rd order amplifier is limited by the generated effective Kerr coupling, we manually add an intrinsic cross Kerr coupling, $k_{ab} \varphi_a^2 \varphi_b^2$, in the SoP-3rd order truncated Lagrangian, and observe the saturation power of the amplifier as we tune $k_{ab}$ (see Fig~\ref{fig:Kerr_compensation}(a) and (b), SoP-3rd line). We observe that in both small $\beta$ (see Fig.~\ref{fig:Kerr_compensation}(a), $\beta = 1.2$) and large $\beta$ (see Fig.~\ref{fig:Kerr_compensation}(b), $\beta = 10.0$), as we tune the intrinsic Kerr term, the saturation power is maximized at the point indicated by the dashed red line. This maximum corresponds to the value of the intrinsic Kerr term that best cancels the generated Kerr coupling ($k_{ab} = -\text{Re}[k_{ab}^{\textrm{eff}}]$) and hence provides a maximum boost to the saturation power. We also notice that the maximum peak on Fig.~\ref{fig:Kerr_compensation}(a) has a shift from the full compensation point ($k_{ab} = -\textrm{Re}[k_{ab}^{\textrm{eff}}]$). This is caused by the existence of imaginary term of $f_\Sigma$. In perturbation analysis, if we turn off the imaginary part of $f_\Sigma$, the peak is perfectly centered at the full-compensation point.

We also compare the saturation power obtained with SoP 3rd order (blue lines) to the StP with intrinsic cross-Kerr term $k_{ab}$ (orange lines). In order to make the comparison more direct, we shift $k_{ab}$ for the StP-Kerr amplifier by the computed value of the generated $\text{Re}[k_{ab}^{\textrm{eff}}]$ of the SoP-3rd order amplifier (i.e. we line up the peaks). We observe that away from the saturation power peak the two models are in good agreement, which supports the correspondence. Further, if we focus on $k_{ab} = 0$ point on the plot, i.e. the point at which SoP-3rd order model has no added intrinsic $k_{ab}$, the saturation power of SoP-3rd order amplifier (blue lines) matches the shifted StP cross-Kerr nonlinear amplifier (orange lines) in both Fig.~\ref{fig:Kerr_compensation}(a) and (b). Therefore, we conclude that it is indeed the generated Kerr coupling that is limiting the saturation power of the SoP-3rd order model. 

However, near the saturation power maximum the two models diverge: the saturation power of the StP-Kerr amplifier becomes infinite as the intrinsic Kerr nonlinearity becomes zero, while the saturation power of the SoP-3rd order amplifier remains finite. This is caused by the imagniary part of $k_{ab}^{\textrm{eff}}$ and the $q_{ab}^{\textrm{eff}}$, which cannot be compensated by a real intrinsic cross-Kerr coupling $k_{ab}$. 

We can understand the $\text{Im} [k_{ab}^{\textrm{eff}}]$ and the $q_{ab}^{\textrm{eff}}$ terms as a  two-photon loss channel, i.e. in which a photon in the signal mode and a photon in the idler mode combine and are lost in the pump mode. Both of the terms can be mapped to an imaginary energy which represents the decay of the signal and idler mode fluxes. Specifically, $\textrm{Im}(f_\Sigma)$ term represents the loss of a photon in signal mode and a photon in idler mode to a pump photon with frequency $\omega = \Sigma$, while $\textrm{Im}(f_\Delta)$ term represents the loss to a $\omega = \Delta$ pump photon.

\subsection{Fifth and Higher Order Nonlinearities} \label{subsec:higher_order}
As we pointed out in Eq.~\eqref{eq:saturation_flux_SoP_3rd_5th}, the saturation flux increases as we decrease the three-mode coupling strength $g$ (by increase $\beta$). However, as we increase $\beta$, the saturation flux diverges from the SoP-3rd order model (see Fig.~\ref{fig:sweep_beta}). This is because the saturation flux is so large that higher order nonlinear couplings becomes the limiting mechanism to the saturation flux. In this subsection, we focus on the higher order couplings and show how they limit the saturation flux of the amplifier.

At the Kerr nulling point, $\varphi_\textrm{ext} = 2\pi$, the Kerr couplings are turned off, and hence the next nonzero order of nonlinear couplings are fifth order in mode fluxes. The fifth order terms in the expansion of the dimensionless potential energy of the JRM, Eq.~\eqref{eq:Edimless}, are
\begin{equation}
    \mathcal{E}_{\textrm{JRM}}^{(5)} = h_{a} \varphi_{a}^{3} \varphi_{b} \varphi_{c} + h_{b} \varphi_{a} \varphi_{b}^{3} \varphi_{c} + h_{c} \varphi_{a} \varphi_{b} \varphi_{c}^{3},
\end{equation}
where $h_{a} = h_{b} = \frac{1}{24 \beta} \sin \left( \frac{\varphi_\textrm{ext}}{4} \right)$ and $h_{c} = \frac{1}{6 \beta} \sin \left( \frac{\varphi_\textrm{ext}}{4} \right)$. To understand the direct effects of the fifth order couplings, we apply stiff-pump approximation and only include 3rd and 5th order nonlinear coupling terms into the EOMs (kerr couplings are turned off at $\varphi_{ext} = 2\pi$). Among the three fifth order terms, $h_a$ and $h_b$ terms are more significant as in stiff-pump approximation where $c$ mode is treated as stiff, the term $h_c \varphi_c^2 \varphi_a \varphi_b \varphi_c$ only shifts the pump mode flux to reach the desired gain $G_0$ and does not causes saturation. However, $h_a \varphi_a^2 \varphi_a \varphi_b \varphi_c$ and $h_b \varphi_b^2 \varphi_a \varphi_b \varphi_c$ terms dynamically shift the effective third order coupling strength as we increase the input signal power, which saturates the amplifier. 

Again, we apply perturbation theory to analyze the StP-5th order amplifier following the discussion in subsection~\ref{subsec:SoP_3rd}. The lowest order solution of the signal and idler mode fluxes are at first order, which repeats the solution of the ideal parametric amplifier. The next nonzero correction appears at third order with equation $[M].[\varphi^{(3)}] = - [i].[\varphi_d^{(3)}]$, where $[i]$ is a $2 \times 2$ diagonal matrix with elements $\{i,-i\}$ and the corresponding drive term is
\begin{widetext}
\begin{equation}
    [\varphi_d^{(3)}] = \left( 
    \begin{array}{c}
    12 h_a \varphi_c \left\vert \varphi_a^{(1)}\right\vert^2 \varphi_{b}^{(1)*} + 6 h_a \varphi_c^* \left( \varphi_a^{(1)} \right)^2 \varphi_b^{(1)} + 6 h_b \varphi_c \left\vert \varphi_{b}^{(1)}\right\vert^2 \varphi_{b}^{(1)*} \\
    6 h_a \varphi_c^{*} \left\vert \varphi_{a}^{(1)}\right\vert^2 \varphi_{a}^{(1)} + 12 h_b \varphi_{c}^{*} \left\vert \varphi_b^{(1)} \right\vert^2 \varphi_a^{(1)} + 6 h_b \varphi_c \left( \varphi_b^{(1)*}\right)^2 \varphi_a^{(1)*}
    \end{array}
        \right).
\end{equation}
\end{widetext}

In the high-gain limit, the third order correction of the signal mode flux is
\begin{equation}
    \varphi_a^{(3)} \sim 4 \frac{h}{g} \left(1+\frac{\tilde{\gamma}_a}{\tilde{\gamma}_b}\right) G_0^2 \varphi_{a,\textrm{in}}^3,
\end{equation}
where $h = h_a = h_b$ is the dimensionless fifth order coupling strength. Following the same method, we get an estimate on the saturation flux
\begin{equation}
    \varphi_{a,\textrm{in},\pm 1\textrm{dB}} \sim \sqrt{\epsilon \frac{g}{4h} \left( 1 + \frac{\tilde{\gamma}_a}{\tilde{\gamma}_b} \right)} G_0^{-3/4}.
    \label{eq:StP_5th_sat_flux}
\end{equation}

We note that the ratio $g/h$ is independent of $\beta$. As we increase $\beta$ to reduce the limitation placed by SoP-3rd order model, Eq.~\eqref{eq:saturation_flux_SoP_3rd_5th}, we eventually hit the limit that is given by StP-5th order nonlinear model, Eq.~\eqref{eq:StP_5th_sat_flux}, i.e. the dominating limiting mechanisms on saturation flux switches.

To be more explicit, similar to the effective cross-Kerr compensation illustrated in subsec.~\ref{subsec:StP_4th}, we add fifth order nonlinear coupling terms into the SoP-3rd order nonlinear model, which is labeled as ``SoP-3rd+5th'' in Fig.~\ref{fig:Kerr_compensation} (green lines). In the small $\beta$ regime [Fig.~\ref{fig:Kerr_compensation}(a)], except around the generated cross-Kerr full compensation region, the SoP-3rd+5th order nonlinear model closely follows the SoP-3rd order model, especially at $k_{ab} = 0$ point where there is no intrinsic $k_{ab}$ added to both of the models. This indicates that at low $\beta$ regime, the dominating limitation on the saturation power is given by the generated effective cross-Kerr coupling from the SoP-3rd order nonlinear coupling. However, when $\beta$ is large [Fig.~\ref{fig:Kerr_compensation}(b)], the saturation flux calculated from these two models disagrees. With additional fifth order nonlinear couplings, the saturation flux is heavily suppressed, which shows that the fifth order nonlinear couplings dominates the SoP-3rd effects in limiting the saturation power of the amplifier. 

Furthermore, in the large $\beta$ regime, the fifth order nonlinear couplings in the JPA Lagrangian is the dominating limitation on the saturation power in full nonlinear EOMs of JPA among all the nonlinear couplings. To prove it, we numerically solve the saturation flux of the StP-5th order truncated model of JPA (dark green solid line in Fig.~\ref{fig:sweep_beta}) and compare it with saturation flux obtained from the full nonlinear EOMs (blue line in Fig.~\ref{fig:sweep_beta}). The saturation flux from StP-5th order model matches the saturation flux of full nonlinear JPA model in large $\beta$ regime perfectly. 

The saturation flux computed by numerical integration of StP-5th order nonlinear model is independent of parameter $\beta$, which agrees with the perturbation analysis. To further validate the perturbation theory, we plot the saturation flux from third order perturbation in Fig.~\ref{fig:sweep_beta}(a) (dark green dashed line) for comparison. We notice that the perturbation result does not have a good quantitative agreement with the numerical solution. This is because the saturation flux is outside the radius of convergence of the perturbation series. If we tighten the creteria for amplifier saturation to the signal mode flux that causes the gain to change by $\pm 0.1$~dB instead, the third order perturbation on StP-5th has much better agreement with the numerical solutions (see Fig.~\ref{fig:sweep_beta}(b) dark green dashed line and solid line). However, to perfectly match the numerical solution, we need next order correction, i.e. fifth order correction to signal mode flux. The result saturation flux is plotted in Fig.~\ref{fig:sweep_beta}(b) as the red dot-dashed line.

Similarly, for the higher order nonlinear couplings in the Lagrangian, e.g. the seventh order in the Hamiltonian, we can still apply the perturbation theory to analyze the saturation flux. Here we focus on one of the seventh order couplings, $-l_{aa} \varphi_{a}^{4} \varphi_{a} \varphi_{b} \varphi_{c}$, to finalize the discussion. According to Eq.~\eqref{eq:H_expansion}, $l_{aa}$ is $\sin \left( \frac{\varphi_\textrm{ext}}{4}\right) / (1920 \beta)$. We still consider the truncated EOMs of the amplifier under stiff-pump approximation.

Following the same procedures discussed above, the lowest order solution of signal and idler mode fluxes are in first order and are given by the ideal parametric amplifier solution in Eq.~\eqref{eq:stpSignalIdler}. However, the next nonzero correction to signal and idler mode fluxes appears at fifth order with the corresponding drive term,
\begin{equation}
    [\varphi_{d}^{(5)}] = 20 \left\vert \varphi_a^{(1)} \right\vert^2 \left( \begin{array}{c}
        3 l_{aa} \left\vert \varphi_a^{(1)} \right\vert^2 \varphi_b^{(1)*} \varphi_c + 2 \left(\varphi_a^{(1)} \right)^2 \varphi_b^{(1)} \varphi_c^* \\ 
        l_{aa} \left\vert \varphi_{a}^{(1)} \right\vert^2 \varphi_a^{(1)} \varphi_c^{*}
    \end{array}
    \right).
\end{equation} 
The saturation flux given by this order of perturbation theory obeys 
\begin{equation}
    \varphi_{a,\textrm{in}, \pm 1 \textrm{dB}} \sim \left(\frac{g}{l_{aa}}\right)^{1/4} G_0^{-5/8}.
\end{equation}
This limit does not depend on $\beta$ either. With StP-7th order truncated nonlinear model, where we include 3rd, 5th and 7th order nonlinear couplings in JPA Lagrangian (even orders are turned off at $\varphi_{\textrm{ext}} = 2\pi$), the existence of the 7th-order nonlinear couplings contributes to a small correction to the saturation flux at large $\beta$. However, the fifth order term remains the dominant factor in determining the saturation flux.
 
To conclude this section, for a JRM based JPA that is operated at the nulling point with fixed mode frequencies and mode linewidth, saturation flux can be increased by increasing $\beta$ which suppresses the effects of generated Kerr couplings. As we move to large $\beta$ regime, if we want to further improve the saturation power of the amplifier, we need to reduce the fifth and higher order nonlinear coupling strengths with respect to the third order coupling strength in the Lagrangian. In Ref.~\cite{GQLiu2017}, we notice that the imperfect participation ratio $p \neq 1$ caused by nonzero linear inductance in series of JRM circuit, is one of the candidates for the suggested suppression, which will be discussed in the following sections.

\section{Effects of Participation Ratio} \label{sec:discussion}

\begin{figure*}[htbp!]
    \centering
    \includegraphics[width = \textwidth]{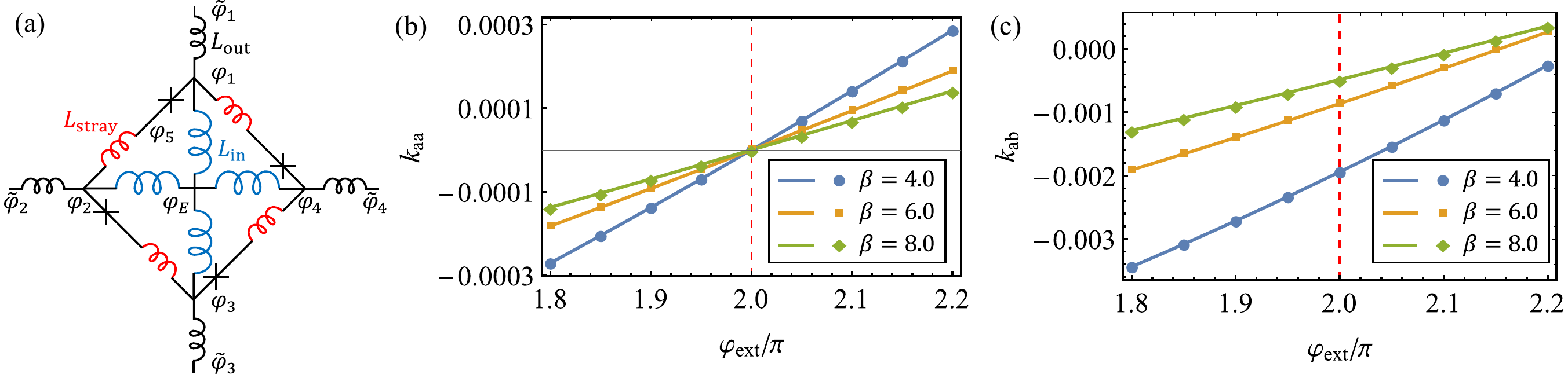}
    \caption{In (a), we show a more realistic circuit model for JRM, in which we include stray inductance $L_{\textrm{stray}}$ in series of Josephson junctions and outer linear inductance $L_{\textrm{out}}$ in series of JRM. The fluxes associated with each nodes are labeled on the plot. In (b) and (c), we calculate the Kerr coupling strength $k_{aa}$ (b) and $k_{ab}$ (c) as we sweep external magnetic field bias $\varphi_{\textrm{ext}}$ when participation ratio $1/p = 1.1$. The perturbation solution (lines) and numerical solution (dots) agree well. In (b) for all three $\beta$ values, the self-Kerr coupling strength $k_{aa}$ can always be turned off at the Kerr nulling point $\varphi_\textrm{ext} = 2\pi$. However, in (c), we notice that the the magnetic field bias $\varphi_{\textrm{ext}}$ to turn off cross-Kerr coupling $k_{ab}$ depends on the choice of $\beta$. This means the exact Kerr nulling point of the does not exist any more when the participation ratio is not unity. Parameters chosen: three mode decay rates are $\gamma/(2\pi) = 0.1$~GHz, the critial current of the junctions is $i_c = 1.0~\mu$A. The outer linear inductance ratio $\zeta=0.1$. The rest of the circuit elements are set by the mode frequencies at $\varphi_\textrm{ext} = 2\pi$, and they remains when we tune the external flux bias.}
    \label{fig:participation}
\end{figure*}

In this section, we focus on the effects of reducing participation ratio by introducing outer linear inductors in series with the JRM circuit [$L_{\textrm{out}}$ in Fig.~\ref{fig:participation}(a)]. 

When there are external resonators connected to the JRM, the flux injected from the microwave ports is shared between the JRM and the external resonators and hence the JRM nonlinearity is attenuated. To model this effect, four outer linear inductors $L_{\textrm{out}}$ are added in series with the JRM circuit [see Fig.~\ref{fig:participation}(a)]. These inductors and the JRM can be treated as a ``flux-divider'' type circuit. Further, as the input-output ports are connected to the outer nodes and there is no capacitors connecting the inner nodes to ground, we treat the fluxes of outer nodes ($\tilde{\varphi}_j$) as free coordinates, while the inner node fluxes ($\varphi_j$) are restricted by the Kirchhoff's current relation. The potential energy of JRM becomes
\begin{equation}
\begin{aligned}
    E & = E_{\textrm{out}} + E_{\textrm{JRM}} \\
    & = \sum_{j} \frac{\phi_0^2}{L_{\textrm{out}}} \left( \tilde{\varphi}_j - \varphi_j\right)^2 + E_{\textrm{JRM}}(\varphi_1,\varphi_2, \varphi_3, \varphi_4).
    \label{eq:participation_E}
\end{aligned}
\end{equation}
The EOM for node flux $\tilde{\varphi}_{j}$ are
\begin{equation}
    \ddot{\tilde{\varphi}}_j + \frac{1}{C_j L_\textrm{out}} \left( \tilde{\varphi}_j - \varphi_j \right) = \text{IN}_j,
    \label{eq:EOM_pnq1}
\end{equation}
where $j = 1,2,3,4$ and the node capacitance $C_j = C_a$ for $j = 1,3$ and $C_j = C_b$ for $j = 2,4$. The right hand side, $\text{IN}_j$ is the corresponding input terms derived in Eq.~\eqref{eq:node_EOM} for each node flux. The inner node fluxes $\varphi_j$ are restricted by 
\begin{widetext}
\begin{equation}
    \tilde{\varphi}_j = \varphi_j + \zeta \left[ \frac{1}{\beta}\sin\left(\varphi_j-\varphi_{j+1}+\frac{\varphi_\textrm{ext}}{4}\right) - \frac{1}{\beta} \sin \left( \varphi_{j-1} - \varphi_j + \frac{\varphi_\textrm{ext}}{4}\right)+ \frac{1}{4}\left(3\varphi_j - \sum_{k\neq j}\varphi_k\right)\right],
    \label{eq:inner_node_constrain}
\end{equation}
\end{widetext}
where $\zeta = L_\textrm{out}/L_\textrm{in}$ and we apply index convention that $\varphi_{0} = \varphi_4$, $\varphi_{5} = \varphi_{1}$. As the symmetry of the JRM still persist, the normal mode profiles in terms of the outer node fluxes $\tilde{\varphi}$'s are identical to the ones without outer linear inductance, i.e. the normal mode coordinates are given by $[\tilde{\varphi}_M] = [\mathcal{A}^{-1}].[\tilde{\varphi}]$, where the model matrix $[\mathcal{A}]$ is identical to Eq.~\eqref{eq:model_matrix}. This can also be derived from the linearization of the JPA's EOMs [Eq.~\eqref{eq:EOM_pnq1}] and the constrains in Eq.~\eqref{eq:inner_node_constrain}. But the frequencies of the normal modes are shifted
\begin{subequations}
\begin{align}
        \omega_{a,(b)}^2  & = \frac{1}{2 C_{a,(b)} L_\textrm{in}} \frac{\beta + 2 \cos \left( \frac{\varphi_\textrm{ext}}{4}\right)}{\beta+\beta \zeta + 2 \zeta \cos \left( \frac{\varphi_\textrm{ext}}{4}\right)}, \\
        \omega_c^2 & = \frac{1}{C_c L_\textrm{in}} \frac{\beta + 4 \cos \left( \frac{\varphi_\textrm{ext}}{4}\right)}{\beta+\beta \zeta + 4 \zeta \cos \left( \frac{\varphi_\textrm{ext}}{4}\right)},
\end{align}
\end{subequations}
where $C_c = \frac{4 C_a C_b}{C_a + C_b}$. 

The question of how the nonlinear couplings shift when we add $L_\text{out}$ into the JRM circuit is hard to directly analyze by the expanding the JRM potential energy in terms of normal modes around the ground state, as the constrains [Eq.~\eqref{eq:inner_node_constrain}] are hard to invert. To obtain the nonlinear coupling strengths, we can either numerically calculate the derivatives of the potential energy with respect to the mode fluxes or using analytical perturbation expansion to get an approximate inversion relation of Eq.~\eqref{eq:inner_node_constrain} and find the non-linear couplings. Here we stop at fourth-order non-linearities (in energy).

To solve the self-Kerr $k_{jj}$ and cross-Kerr $k_{ij}$ nonlinear coupling strengths, we can calculate the fourth order derivatives of the circuit potential energy $E$ with respect to the normal coordinates $\tilde{\varphi}_a$, $\tilde{\varphi}_b$ and $\tilde{\varphi}_c$, i.e. 
\begin{equation}
    k_{jj} = \frac{1}{24} \frac{\partial^4 \mathcal{E}}{\partial\tilde{\varphi}_{j}^4} , \, 
    k_{ij} = \frac{1}{4} \frac{\partial^4 \mathcal{E}}{\partial\tilde{\varphi}_{i}^2 \partial\tilde{\varphi}_{j}^2} .
    \label{eq:k_derivative}
\end{equation}
where $\mathcal{E}$ is dimensionless energy of JRM circuit defined as $\mathcal{E} = (L_\textrm{in}/\phi_0^2) E$.

It is straightforward to use inner node fluxes to express the energy $E$ in Eq.~\eqref{eq:participation_E}, and hence find an analytical expression for the derivatives with respect to inner node fluxes.  However, to calculate derivatives with respect to the outer node fluxes requires the Jacobian matrix $[J] = [\frac{\partial \varphi}{\partial \tilde{\varphi}}]$, which effectively requires inversion of the constrains in Eq.~\eqref{eq:inner_node_constrain}.

To analytically solve this problem and give us a hint on the how the outer linear inductance will affect the nonlinear couplings, we apply the perturbation expansion around the ground state ($\tilde{\varphi}_j = 0$) to obtain an approximate inverse transformation and find the Jacobian. To simplify the discussion, we assume $C_a = C_b$. We note that this assumption does not affect the nonlinear coupling strengths which are independent of the $c$ mode. Further, the method we discussed below can be easily generalized to the case when $C_a \neq C_b$. 

We at first define a set of new variables using the normal mode transformation matrix $[\mathcal{A}]$, but use the inner node fluxes instead, noted as $[\varphi_M] = [\mathcal{A}^{-1}].[\varphi]$. Therefore, the relation in Eq.~\eqref{eq:inner_node_constrain} using normal coordinates $[\tilde{\varphi}_M]$ and inner node coordinates $[\varphi_M]$ is
\begin{equation}
    \tilde{\varphi}_j = (1+\zeta) \varphi_j + \mathit{(2)} \frac{2 \zeta}{\beta} \frac{\partial}{\partial \varphi_j}\mathcal{E}_{\textrm{JRM}}^{(0)}
    \label{eq:constrain_mode}
\end{equation}
where $\mathcal{E}_{\textrm{JRM}}^{(0)}$ is given in Eq.~\eqref{eq:Edimless}, the factor $\mathit{(2)}$ only exists for $a$ and $b$ modes.
Here we only focus on the three nontrivial modes, $\tilde{\varphi}_a$, $\tilde{\varphi}_b$ and $\tilde{\varphi}_c$. 
The circuit ground state is assumed to be stable and at $\tilde{\varphi}_a = \tilde{\varphi}_b = \tilde{\varphi}_c = 0$ (which we confirm numerically). Further, at this stable ground state, the inner node fluxes are also zero. Since we only focus on the Kerr coupling strength in the vicinity of the ground state, the exact inner node fluxes that obey the inverse relation of Eq.~\eqref{eq:constrain_mode} can be expanded in series of the small oscillations of the normal modes $\tilde{\varphi}_j$'s. That is, $\varphi_j \sim 0 + \varphi_j^{(1)} + \varphi_j^{(2)} + ...$.

We plug the expansion of inner node fluxes back to Eq.~\eqref{eq:inner_node_constrain} and match the terms with order-by-order. The lowest order solutions appear at the first order in normal coordinates 
\begin{subequations}
\begin{align}
    \varphi_{a,b}^{(1)} & = \left[1+\zeta+ \frac{2\zeta}{\beta}\cos\left(\frac{\varphi_\textrm{ext}}{4}\right)\right]^{-1}\tilde{\varphi}_{a,b}, \\
    \varphi_c^{(1)} & = \left[1+\zeta+ \frac{4\zeta}{\beta}\cos\left(\frac{\varphi_\textrm{ext}}{4}\right)\right]^{-1} \tilde{\varphi}_c.
\end{align}
\end{subequations}
At this order, we can extract the definition of participation ratio for signal and idler mode as $p_{a,b} = \left[1+\zeta+ \frac{2\zeta}{\beta}\cos\left(\frac{\varphi_\textrm{ext}}{4}\right)\right]^{-1}$ and for pump mode as $p_c = \left[1+\zeta+ \frac{4\zeta}{\beta}\cos\left(\frac{\varphi_\textrm{ext}}{4}\right)\right]^{-1}$. If we bias the circuit at $\varphi_{\textrm{ext}} = 2\pi$, all three participation ratios become $p_0 = \frac{1}{1+\zeta}$. 

The second order correction to the inner node fluxes are
\begin{subequations}
\begin{align}
    \varphi_a^{(2)} & = p_a \frac{2\zeta}{\beta}\sin\left(\frac{\varphi_\textrm{ext}}{4}\right) \varphi_{b}^{(1)} \varphi_{c}^{(1)}, \\
    \varphi_b^{(2)} & = p_b \frac{2\zeta}{\beta}\sin\left(\frac{\varphi_\textrm{ext}}{4}\right) \varphi_{b}^{(1)} \varphi_{c}^{(1)}, \\
    \varphi_c^{(2)} & = p_c \frac{\zeta}{\beta}\sin\left(\frac{\varphi_\textrm{ext}}{4}\right) \varphi_{a}^{(1)} \varphi_{b}^{(1)}. 
\end{align}
\end{subequations}
The corresponding approximate inverse transformation of Eq.~\eqref{eq:inner_node_constrain} is $\varphi_j \sim \varphi_j^{(1)}(\{\tilde{\varphi}\}) + \varphi_j^{(2)}(\{ \tilde{\varphi} \})$ for $j=a,b,c$. At second order, it is sufficient to calculate the three mode coupling strength, as we only need at most the second order derivatives to the Jacobian matrix elements. The dimensionless three mode coupling strength are
\begin{equation}
    g(\zeta) \equiv \frac{\partial^3 \mathcal{E}}{\partial \tilde{\varphi}_a \partial \tilde{\varphi}_b \partial \tilde{\varphi}_c} = p_a p_b p_c g(0),
    \label{eq:g_pneq1}
\end{equation}
where $g(0)$ is the three-mode coupling strength with unit participation ratio. Based on Eq.~\eqref{eq:g_pneq1}, decreasing the participation ratio by increasing $L_{\textrm{out}}$ reduces the corresponding third order coupling strength, which is beneficial to reduce the limitation placed by the effective cross-Kerr nonlinearity generated by SoP-3rd order nonlinear couplings, and hence it is beneficial to improving the saturation power of the amplifier in the small $\beta$ regime.

However, to calculate the fourth order derivatives, we need at least third order correction to the inner node fluxes. Following the same strategy, the third order correction of the inverse transformation for normal coordinate $\varphi_a$ is
\begin{widetext}
\begin{equation}
    \begin{aligned}
     \varphi_a^{(3)} = \frac{p_a^4 \zeta}{12 \beta} \cos\left(\frac{\varphi_\textrm{ext}}{4}\right)\tilde{\varphi}_a^3 & + 
        \frac{p_a^2 p_b^2 \zeta}{4 \beta^2} \left[
        \beta \cos\left(\frac{\varphi_\textrm{ext}}{4}\right) + 8 p_c \zeta \sin^2 \left(\frac{\varphi_\textrm{ext}}{4}\right)
        \right] \tilde{\varphi}_a \tilde{\varphi}_b^2 \\
        & + 
        \frac{p_a^2 p_c^2 \zeta}{4 \beta^2} \left[
        \beta \cos\left(\frac{\varphi_\textrm{ext}}{4}\right) + 4 p_b \zeta \sin^2     \left(\frac{\varphi_\textrm{ext}}{4}\right)
        \right] \tilde{\varphi}_a \tilde{\varphi}_c^2
    \end{aligned}
\end{equation}
\end{widetext}
and the relations for $\varphi_b^{(3)}$ and $\varphi_c^{(3)}$ can be derived similarly. The inverse relation from Eq.~\eqref{eq:inner_node_constrain} is $\varphi_j \sim \varphi_j^{(1)}(\{\tilde{\varphi}\}) + \varphi_j^{(2)}(\{ \tilde{\varphi} \})  + \varphi_j^{(3)}(\{ \tilde{\varphi} \})$. The Kerr coupling strengths can be obtained from Eq.~\eqref{eq:k_derivative} with Jacobian derived from the perturbation expansion. For example $k_{ab}$ is 
\begin{widetext}
\begin{equation}
    k_{ab}(\zeta) = -\frac{
    \beta^3 \left\lbrace \beta (1+\zeta)\cos\left(\frac{\varphi_\textrm{ext}}{4}\right) + 2 \zeta \left[
    -3 + \cos\left( \frac{\varphi_\textrm{ext}}{4}\right) + 8 \sin\left( \frac{\varphi_\textrm{ext}}{4}\right)
    \right]
    \right\rbrace
    }{ 16
    \left[\beta +\beta \zeta + 2 \zeta \cos\left( \frac{\varphi_\textrm{ext}}{4}\right)
    \right]^4
    \left[\beta+\beta\zeta + 4 \zeta \cos\left( \frac{\varphi_\textrm{ext}}{4}\right)
    \right]
    }.
    \label{eq:perturbed_kab}
\end{equation}
\end{widetext}

The self-Kerr coupling strength $k_{aa}$ and the cross-Kerr coupling strength $k_{ab}$ are plotted in Fig.~\ref{fig:participation}(b) and (c), respectively. The Kerr nonlinear coupling strengths ($k_{aa}$ and $k_{ab}$) are calculated via both numerical method (dots) and the above perturbation method (lines). In all three $\beta$ values, the perturbation analysis matches the numerical solution well. Further, we notice that the self-Kerr coupling strength can still be turned off at the $\varphi_\textrm{ext} = 2 \pi$ (Kerr nulling point) no matter what $\beta$ value we choose [see Fig.~\ref{fig:participation}(b)]. But the cross-Kerr couplings cannot be turned off at this magnetic bias point when participation ratio is not unity [see Fig.~\ref{fig:participation}(c)]. 

The breakdown of the universal Kerr nulling point is also demonstrated by Eq.~\eqref{eq:perturbed_kab}. The $\varphi_{\textrm{ext}}$ that makes the numerator of Eq.~\eqref{eq:perturbed_kab} zero depends on the choice of $\beta$ and $\zeta$. This indicates that as we turn the participation ratio to be smaller than unity, some nonlinear couplings that are previously killed by Kerr nulling point can reappear in the JPA Lagrangian. These extra nonlinear couplings are a consequence of the nonlinearity of the inner JRM circuit. As we mentioned, the JRM circuit with outer linear inductance shown in Fig.~\ref{fig:participation}(a) can be treated as a phase divider, i.e. the phase across the outer nodes are divided to the phase across the outer linear inductors ($L_{\textrm{out}}$) and the phase across inner JRM nodes governed by the effective inductance of the inner JRM. Naively, if the divider is linear, we would expect the JRM with outer linear inductance generates nonlinear coupling strengths that are suppressed by the participation ratio (which does not depend on the mode flux), e.g. $k_{ab} (\zeta) = p_a^2 p_b^2 k_{ab} (0)$ and $k_{ab}(0) = 0$ is the cross-Kerr coupling strength of a JRM without outer linear inductance. However, as the effective inductance of inner JRM circuit is nonlinear, the total phase is not divided linearly, i.e. the more precise participation ratio defined as $p = \tilde{\varphi} / \varphi$ will change as the input flux oscillates as it is indeed a function of the outer node fluxes. Therefore, the normal modes experience extra nonlinearities as compared to the naive analysis. The re-apprearace of these extra nonlinearities will limit the saturation power of the amplifier.

However, for a general $\varphi_{\text{ext}}$, the Kerr couplings are suppressed roughly by $\sim p^4$. If we calculate one order up, the fifth order nonlinear coupling strength is suppressed by $\sim p^5$. This indicates that the non-unity participation ratio can help to suppress the higher order nonlinear couplings with respect to the third order, which is beneficial for improving the saturation power of the amplifier. We will focus on the quantitative understand of how these two factors compete with each other and further optimize the saturation power of the amplifier in next section. 

\section{Optimizing the JPA using participation ratio} \label{sec:full_optimization}

As demonstrated in the above section, the outer linear inductance impacts the saturation power of the JPA in both negative and positive ways. In this section we describe the effects of the outer linear inductance quantitatively using numerics to obtain the saturation power of the JPA as we sweep the JRM inductance ratio ($\beta$) and participation ratio ($p$).

\begin{figure}[bp]
    \centering
    \includegraphics[width = 3.2 in]{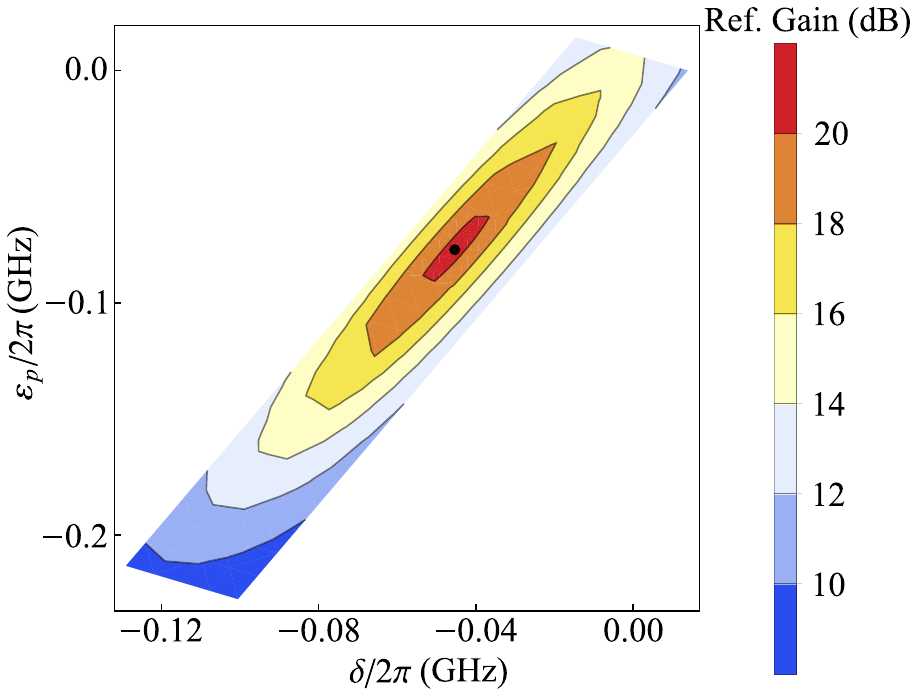}
    \caption{The optimization of the pump configuration. We sweep the signal mode detuning $\delta = \omega_S - \bar{\omega}_a$ and the pump tone detuning $\epsilon_p = \omega_P - (\bar{\omega}_a + \bar{\omega}_b)$ and fix the pump tone strength. The maximum gain is labeled by the black dot. The maximum gain is achieved when the signal tone matches the mode frequency and the pump mode matches the sum frequency of the signal and idler mode. The parameters used: $\gamma/2\pi = 0.2$~GHz, $\bar{\omega}_a/2\pi = 7.5$~GHz, $\bar{\omega}_b/2\pi = 5.0$~GHz, $\beta = 3.0$, $1/p = 8.0$.}
    \label{fig:optimization_freq}
\end{figure}

Because of the presence of the outer linear inductance, even order nonlinear coupling terms reappears in the EOMs. The presence of these higher order couplings results in a shift of the mode frequencies. For example, the nonzero cross-Kerr coupling strength $k_{ac}\varphi_a^2 \varphi_c^2$ and $k_{bc} \varphi_b^2 \varphi_c^2$ causes the signal and idler mode frequencies to be dependent on the pump mode strength, which shifts the signal and idler mode frequencies away from the bare mode frequencies calculated from the normal mode analysis. To correctly pump the amplifier with the sum frequency of mode and idler mode frequencies and probe the signal with the correct signal mode frequency, as well as set the amplifier's small-signal reflection gain to $20$~dB, we need to adjust the pump tone frequency and pump tone strength at the same time. Before we perform the numerical calculation of the amplifier's reflection gain as we tune the input tone strength and extract the saturation power, we need to find  the correct pump configurations and the signal mode frequency under that pump configuration.

To compensate for the frequency shifts and find the optimum pump configuration and corresponding signal mode frequency for JPA, we numerically optimize the pump tone frequency and strength. To solve this optimization problem, we notice that the amplifier is expected to consume the least pump tone input flux to reach the desired small-signal reflection gain when the amplifier is perfectly on resonance with its mode frequencies, i.e. $\omega_S = \omega_a$ and $\omega_P = \omega_a + \omega_b$. Therefore, we split the optimization process into two optimization tasks: (1) for a given input pump tone strength $\varphi_{c,\textrm{in}}$, find the optimimal pump tone frequency and signal mode frequency and (2) find the desired pump tone strength $\varphi_{c,\textrm{in}}$ to get $20$~dB small-signal reflection gain with the corresponding optimized pump tone frequency. In (1), we fix the pump tone strength $\varphi_{c,\textrm{in}}$ and sweep signal tone and pump tone frequencies to find the parameters which maximize the reflection gain (a typical sweep is shown in Fig.~\ref{fig:optimization_freq}). In (2), we use a binary search to find the desired pump strength $\varphi_{c,\textrm{in}}$ for $20$~dB reflection gain. 

\begin{figure}[tbp]
    \centering
    \includegraphics[width = 3.0 in]{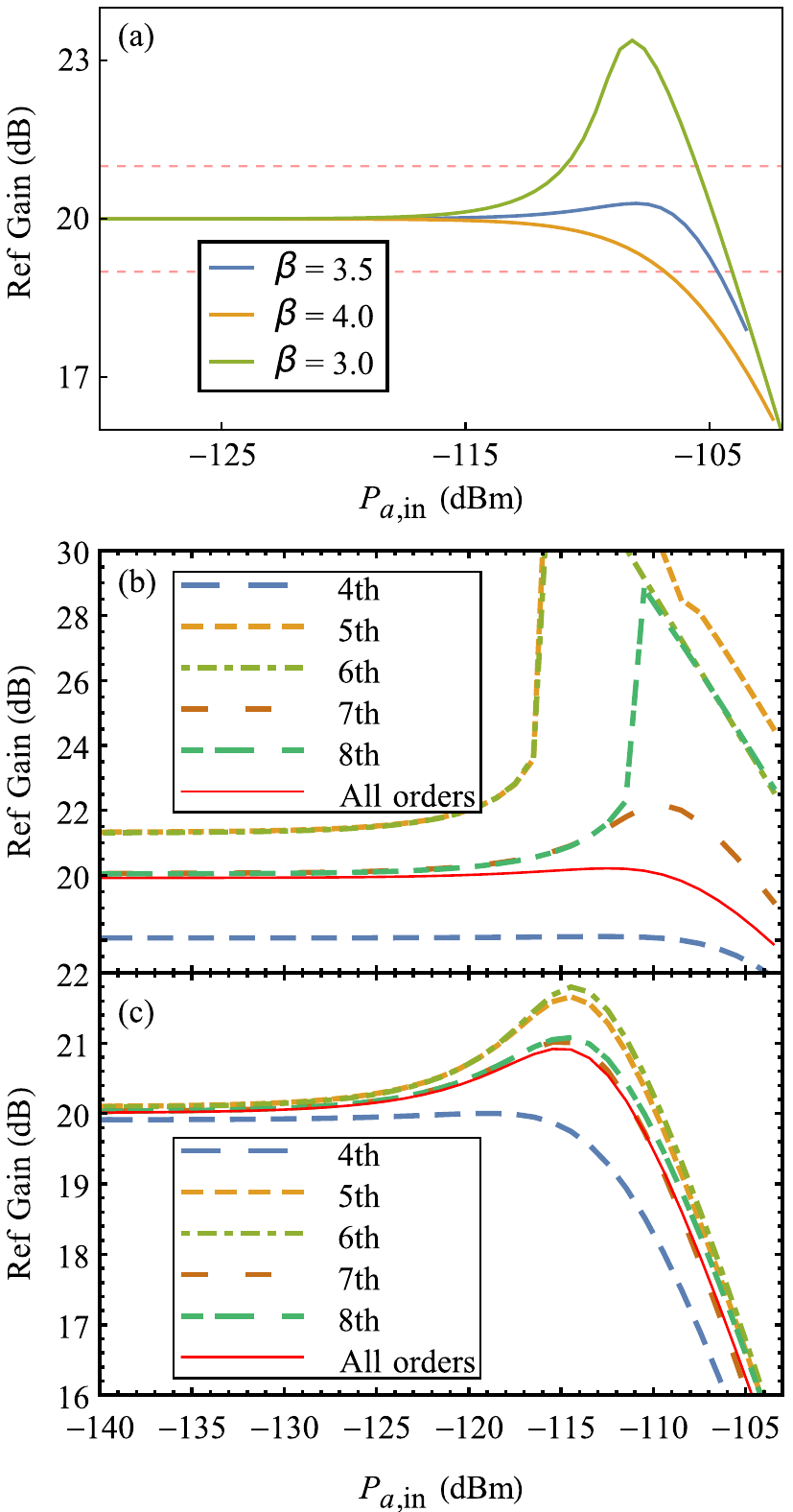}
    \caption{In this plot, we show the reflection gain of the amplifier as we increase the input signal power. We focus on a point which is away from the boundary shown in Fig.~\ref{fig:participation}(a). We test the reflection gain of the truncated model for $\beta = 4.5$ and $1/p \sim 4.0$ as we increase the input signal power $P_{a,\textrm{in}}$ in (c). In the calculation for the truncated model, we only truncate the JRM potential energy to the desired order, but fix the pump configuration as the full-order case. The reflection gain solved from truncated model also converge to the full-order analysis (red solid line) pretty well we truncated to 7th order. But as we decrease $p$ further to push the configuration closer to the boundary ($1/p = 7.0$), the higher and higher order terms are needed to have a good approximation to the full-order performance.}
    \label{fig:convergence}
\end{figure}

The resulting saturation power sweep of the JPA is shown in Fig.~\ref{fig:result}(a). In the large $\beta$ regime ($\beta > 4.0$), as we decrease the participation ratio, the saturation power increases. However, at the same time, the pump power for $20$~dB reflection gain also increase, until the JRM reaches the full nonlinear regime and we cannot inject enough power to get $20$~dB reflection gain anymore. However, in the low $\beta$ regime ($\beta < 4.0$), when the participation ratio is less than unity, even though we firstly optimize the pump configuration to compensate for mode shifting, we still found that the reflection gain of the amplifier increases before it starts to drop (``shark fin''). This causes the amplifier to saturate as gain increases to $21$~dB. If we move out of this regime by reducing the participation ratio or increase $\beta$, the ``shark fin'' reduces and we find a band of sweet spots of the JPA saturation power. The reflection gain of the JPA with configurations around one of the sweet spots is shown in Fig.~\ref{fig:convergence}(a), with the the blue curve corresponding to the sweet spot at $\beta = 3.5$, $1/p = 7.0$. As we decrease $\beta$ to $3.0$, the JPA saturates as gain touches  $21$~dB [green curve in Fig.~\ref{fig:convergence}(a)], while as we increase $\beta$ to $4.0$ the  ``shark fin'' disappears but the saturation power decreases.

To understand the dominating limitations placed by different nonlinear terms in the JPA Hamiltonian, especially around the sweet spot, we truncate the Hamiltonian order-by-order and analyze the performance of the truncated model. We keep the pump configurations identical to the full-order analysis and increase the truncation order from 3rd order to 8th order. In Fig.~\ref{fig:convergence}(b), we focus on the sweet spot $\beta = 3.5$, $1/p = 7.0$, and compare the truncated theory with the full-nonlinear solution. At small signal input, the nonlinear couplings up to 7th order are needed to converge to desired $20$~dB reflection gain. This is a sign that the high order nonlinear coupling terms play an important role in the dynamics of the JPA. As we increase the signal power, the truncation to 4th order analysis does not show an obvious ``shark fin'' feature. However, when we include the higher order coupling terms, e.g. 5th to 8th, the ``shark fin'' appears. The truncated 5th order analysis supports another mechanism that causes the amplifier to saturate to $21$~dB which is different from the one discussion in Ref.~\cite{GQLiu2017}, that is the fifth order terms, e.g. $\varphi_a^2 \varphi_a \varphi_b \varphi_c$ term, can shift the bias condition by shifting the effective third order coupling strength to drive the amplifier towards the unstable regime causing the reflection gain to rise. Further, as we discussed above, the external linear inductors breakdown the perfect nulling point for even order nonlinear couplings, the 6th order and 8th order terms can survive at the nulling point. From 5th order to 8th order truncation, the large signal input behavior oscillates, which is a sign that we are reaching the convergence point of the series expansion casued by the competition between  different orders. We also compare it with a point away from the sweet spot in Fig.~\ref{fig:convergence}(c) ($\beta = 3.5$, $1/p = 4.0$). At this point, the 5th order truncation already converges to $20$~dB reflection gain and the 7th order theory gives a good approximation to full order analysis with moderate input signal power. We conclude that the boost in performance of the amplifier at the sweet spot is a result of taking advantage of all orders, and hence cannot be modeled using a low order truncated theory. 

\section{Effects of tuning the external magnetic field, decay rates, and stray inductance} \label{sec:exp_imperfect}

\begin{figure*}[tbp]
    \centering
    \includegraphics[width = 6.0 in]{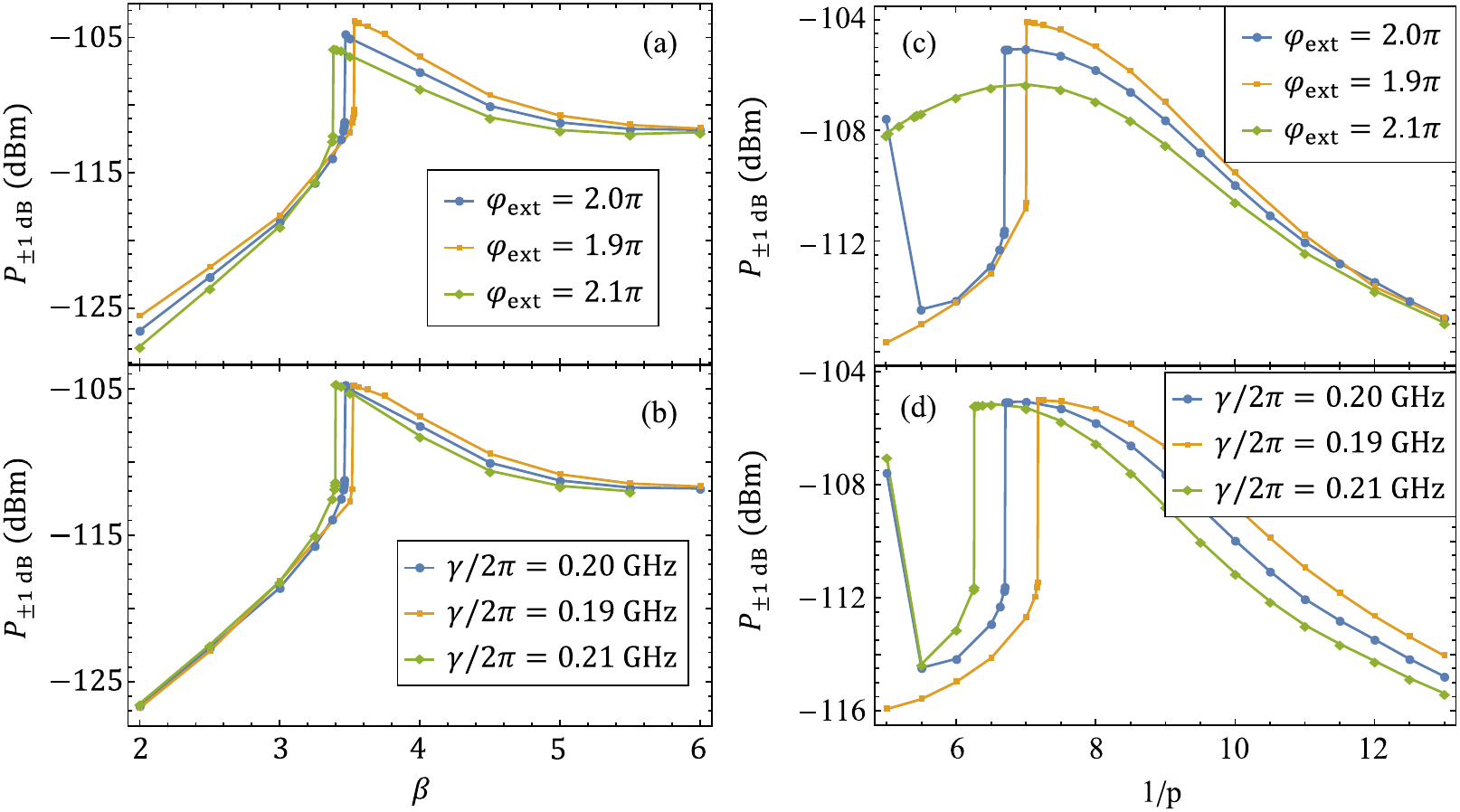}
    \caption{The saturation power of the JPA with external magnetic field bias and mode decay rates perturbation. In (a) and (c) we perturb the JPA external magnetic field bias from $2\pi$ by $\pm 0.1 \pi$. We assume the JPA circuit parameters are fixed with bias $\varphi_{\textrm{ext}} = 2\pi$, and then we operate the JPA at the perturbed magnetic field bias. In (b) and (d) we set the circuit parameters of JPA to change the modes' decay rates from $\gamma/2\pi = 0.2$~GHz by $\pm 10$~MHz. In (a) and (b), we focus on the JPA settings with $1/p = 7.0$ and investigate the effect of the perturbation while in (c) and (d), we focus on the settings with $\beta = 3.5$. }
    \label{fig:perturbation_gamma_phi}
\end{figure*}

In this section, we further explore how the saturation power of the amplifier is affected by the magnetic field bias ($\varphi_{\textrm{ext}}$), the modes' decay rates ($\gamma$), and stray inductance in the JRM loop [$L_{\textrm{stray}}$ in Fig.~\ref{fig:participation}(a)]. 

In Fig.~\ref{fig:perturbation_gamma_phi}, we plots the saturation power of the amplifier as we perturb the magnetic field bias and decay rates of JPA. Here we focus on the line of $1/p=7.0$ in Fig.~\ref{fig:perturbation_gamma_phi}(a) and (b), and focus on the line of $\beta = 3.5$ in (c) and (d). In Fig.~\ref{fig:perturbation_gamma_phi}(a) and (c), we explore the effects of tuning the magnetic field bias. We at first set the JPA circuit parameters at $\varphi_{\textrm{ext}} = 2\pi$. We then operate the JPA at $\varphi_{\textrm{ext}} = 1.9 \pi$ and  $\varphi_{\textrm{ext}} = 2.1 \pi$, respectively. We notice that as we perturb the magnetic field to $\varphi_{\textrm{ext}} = 1.9 \pi$, the optimum saturation power is achieved at larger $\beta$ values [see Fig.~\ref{fig:perturbation_gamma_phi}(a)] and smaller participation ratio $p$ [see Fig.~\ref{fig:perturbation_gamma_phi}(c)]. By tuning $\beta$, the saturation power of the amplifier improves from $-104.8$~dBm to $-103.9$~dBm, while by tuning $p$, it improves to $-104.1$~dBm. This indicates that the optimal magnetic field bias occurs at somewhat lower magnetic field as compared to the Kerr nulling point. The corresponding sweet spot of the amplifier has larger $\beta$ and lower $p$ compare to the present setting. 

In Fig.~\ref{fig:perturbation_gamma_phi}(b) and (d), we change the JPA modes' decay rates by $10$~MHz to explore the effects of different decay rates to the JPA saturation power. In large $\beta$ regime, increasing the JPA mode decay rates causes the regime in which we cannot obtain 20~dB (see Fig.~\ref{fig:result}) gain to become larger. For example at $\gamma/2\pi = 0.21$~GHz, the JPA with $\beta = 6.0$ and $1/p = 7.0$ can no longer reach $20$~dB reflection gain while a comparable JRM with $\gamma/2\pi = 0.20$~GHz could. The amplifier's optimum saturation power is also achieved at a lower $\beta$ value as we increase the decay rates [see Fig.~\ref{fig:perturbation_gamma_phi}(b)]. However, as we tune the decay rates by $\pm 10$~MHz, the maximum saturation power of the amplifier at $1/p = 7.0$ shows little change. Similarly, in Fig.~\ref{fig:perturbation_gamma_phi}(d), we perturb the modes' decay rates by $\pm 10$~MHz on JPA with different $p$ but a fixed $\beta$ ($\beta = 3.5$). The amplifier's optimum saturation power is achieved at a lower $p$ value as we decrease the decay rates [see Fig.~\ref{fig:perturbation_gamma_phi}(d)]], while the maximum saturation power of the amplifier still shows little change.

\begin{figure}[tbp]
    \centering
    \includegraphics[width = 3.2 in]{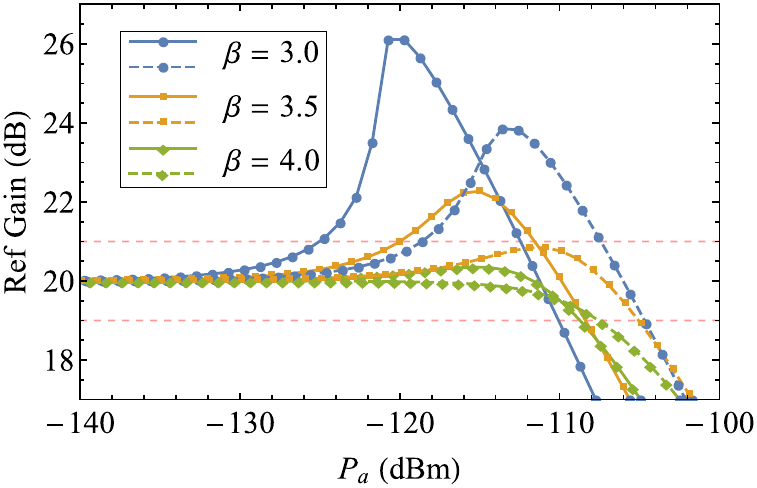}
    \caption{We compare the saturation power of the amplifier without stray inductance ($\alpha = 0$, dashed curves) and with stray inductance ($\alpha = 0.1$ slide curves). We tested three different settings of JPA, $\beta = 3.0$, $3.5$ and $4.0$, respectively. All of them have $1/p = 7.0$. We compare the reflection gain of the amplifier as we increase the signal power $P_a$. For all three cases, the saturation power is suppressed. The existence of the stray inductance enhances the shark fin, which causes the amplifier at previous sweet spot ($1/p = 7.0$ and $\beta = 3.5$) saturates to $21$~dB instead.}
    \label{fig:non_zero_alpha}
\end{figure}

Finally we consider the effect of stray inductors ($L_\textrm{stray}$ in Fig.~\ref{fig:participation}). We include stray inductance such that $\alpha = L_{\textrm{stray}}/ L_{J} = 0.1$ and compare the reflection gain of the amplifier as we increase the signal power ($P_a$). Note that when the stray inductance is nonzero, the Kerr nulling point is shifted away from $\varphi_{\textrm{ext}} = 2\pi$ (see discussion in subsec.~\ref{subsec:L_stray}), especially, when $\alpha = 0.1$, the Kerr nulling point is at $\varphi_{\textrm{ext}} \sim 2.49 \pi$. We will operator the JPA at this magnetic field bias when the participation ratio is not unity. In Fig.~\ref{fig:non_zero_alpha}, we compare three different settings of JPA, $1/p = 7.0, \beta = 3.0$ (blue curves), $1/p = 7.0 , \beta = 3.5$ (orange curves) and $1/p = 7.0, \beta = 4.0$ (green curves). In all three different sittings, we notice enhancement of the ``shark fin'', which causes the JPA at the previous sweet spot ($\beta = 3.5$, orange dashed curve) saturates to $21$~dB instead, which greatly reduce the saturation power at this point (from $-104.8$~dBm to $-120$~dBm). At $\beta = 4.0$, without stray inductors, the reflection gain of the amplifier monotonically decreases as the signal power increases (dashed green line), while at $\alpha = 0.1$ there is a shallow increases (see solid green line). Besides, the saturation power slightly drops from $-107.5$~dBm to $-108.7$~dBm.

\section{Summary and Outlook} \label{sec:summary}
In conclusion, we have investigated the nonlinear couplings of the JRM based JPA and how these different nonlinear couplings controls the performance of the parametric amplifier. In our analysis, we have adapted both perturbative and time-domain numerical methods to give us a full understanding of the circuit dynamics. By considering the full nonlinear Hamiltonian of the device, we show that we can fully optimize the performance of the amplifier, and achieve a $\sim 10$ to $15$~dB improvement of the saturation power of the JRM based JPA for a range of circuit parameters.  Our method for numerically modeling multi-port circuits of inductors, capacitors, and Josephson junctions is also applicable to more complex circuits and pumping schemes, which can create JPAs with addition virtues such as extremely broad (and gain-independent) bandwidth and directional amplification~\cite{Ranzani2015, Aumentado2010, Sliwa2015, Metelmann2014, Metelmann2015, TC-Chien2019}.

\section{Acknowledgement}
The authors gratefully acknowledge fruitful discussions with J. Aumentado, S. Khan, A. Metelmann, and H. T\"{u}reci. C. Liu acknowledges support from the Dietrich School of Arts and Sciences fellowship, and  T.-C. Chien acknowledges support from the Pittsburgh Quantum Institute fellowship. Research was sponsored by the Army Research Office and was accomplished under Grant Number W911NF-18-1-0144.  The views and conclusions contained in this document are those of the authors and should not be interpreted as representing the official policies, either expressed or implied, of the Army Research Office or the U.S. Government. The U.S. Government is authorized to reproduce and distribute reprints for Government purposes notwithstanding any copyright notation herein.   

\section{Appendix}

\subsection{The effect of stray inductance with unit participation ratio} \label{subsec:L_stray}

\begin{figure}[htbp!]
    \centering
    \includegraphics[width = 3 in]{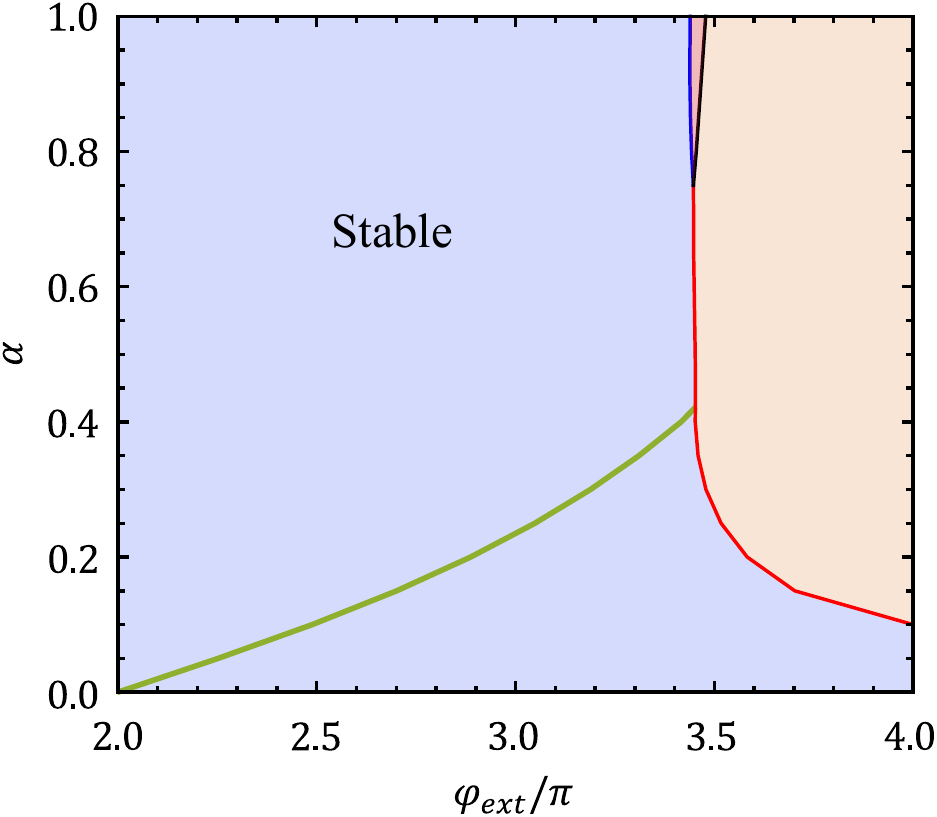}
    \caption{The stability diagram of the ground state of the JPA when we have nonzero stray inductance. We assume $L_\textrm{out}=0$ and set $\beta = 4.0$. The blue region shows the stable region of the JPA ground state, while in orange region, the JPA ground state is doubly degenerate. In the red region, JPA has four-fold degenerate ground state. The green line shows the position of the nulling point.}
    \label{fig:alpha_phase}
\end{figure}

In this section, we focus on the effect of the existence of nonzero stray inductance with unit participation ratio. This discussion is also provided in Ref.~\cite{TC-Chien2019}.

The circuit model of JRM circuit with stray inductance is in Fig~\ref{fig:participation}(a). When the stray inductance is nonzero, similar to shunted JRM circuit, we can write the potential energy of JRM circuit as,
\begin{equation}
    E_{\textrm{JRM}} = \frac{\phi_0^2}{2 L_\textrm{in}} \sum_j \left( \varphi_j -\varphi_E \right)^2 + \sum_j E_{\textrm{arm}}\left( \delta_j \right),
\end{equation}
where $\varphi_E = \frac{1}{4}(\varphi_1 + \varphi_2 + \varphi_3 + \varphi_4)$, the arm energy, $E_{\textrm{arm}}$, is the total energy of the stray inductor and the Josephson junction on one arm of the JRM, $\delta_j = \varphi_j - \varphi_{j+1} + \frac{\varphi_\textrm{ext}}{4}$ is the total phase difference across the $j$-th arm. Take one of the arms as an example, 
\begin{equation}
    H_{\textrm{arm}}(\delta_1) = \frac{\phi_0^2}{2 L_\textrm{stray}} \left(\varphi_1 - \varphi_5\right)^2 - \frac{\phi_0^2}{L_\textrm{J}} \cos \left( \varphi_5 - \varphi_2 + \frac{\varphi_{\textrm{ext}}}{4}\right),
\end{equation}
where the phase on node $\varphi_{5}$ is constrained by the current relation at the corresponding node,
\begin{equation}
    \delta_1 - \Delta \varphi = \alpha \sin \left( \Delta \varphi \right),
        \label{eq:arm_phase}
\end{equation}
where $\alpha = L_\textrm{stray}/L_\textrm{J}$, $\delta_1 = \varphi_1 -\varphi_2 + \frac{\varphi_\textrm{ext}}{4}$ is the total phase difference of the arm and $\Delta \varphi$ is the phase across the junction, defined as $\Delta \varphi = \varphi_5-\varphi_2 + \frac{\varphi_\textrm{ext}}{4}$. Suppose we focus on the case where the external magnetic flux is around $2\pi$, when $\alpha$ is small ($\alpha < 2.80$), the nonlinear relation in Eq.~\eqref{eq:arm_phase} only has a single root when the total phase across the arm is determined.

To determine the self-Kerr $k_{jj}$ and cross-Kerr $k_{ij}$ coupling strengths, we can use the derivatives of the dimensionless JRM energy as,
\begin{equation}
    k_{jj} =  \frac{1}{24} \frac{\partial^4 \mathcal{E}_{\textrm{JRM}}}{\partial\varphi_{j}^4} , \, 
    k_{ij} = \frac{1}{4} \frac{\partial^4 \mathcal{E}_{\textrm{JRM}}}{\partial\varphi_{i}^2 \partial\varphi_{j}^2} .
    \label{eq:k_derivative_alpha}
\end{equation}
Before we carry on the derivative, we appreciate the fact that the phase difference across the arms are linearly dependent on the node fluxes, and the node fluxes are linearly dependent on the normal mode coordinates. Since the inner linear inductance only contribute the energy which are quadratic to the node phases, there will be no contribution to the Kerr couplings. Because the four arms of the JRM is symmetric, the arm Hamiltonian for four arms should have identical form in terms of the phase difference $\delta$. To finalize the calculation, the forth order derivatives with respect to normal modes in general can be calculated as,
\begin{equation}
\begin{aligned}
    \frac{\partial^4}{\partial \varphi_i^2 \partial \varphi_j^2} &  \mathcal{E}_\textrm{JRM}  = \sum_l \frac{\partial^4}{\partial \varphi_i^2 \partial \varphi_j^2} \mathcal{E}_{\textrm{arm}}(\delta_l) \\
    = \sum_l & \left(\frac{\partial^4}{\partial \delta^4} \mathcal{E}_{\textrm{arm}}\right) \left( \frac{\partial \delta_l}{\partial \varphi_i}\right)^2 \left( \frac{\partial \delta_l}{\partial \varphi_j}\right)^2.
    \end{aligned}
\end{equation}
Therefore, for both self-Kerr couplings and cross-Kerr couplings, there is a common factor $\partial_\delta^4 \mathcal{E}_{\textrm{arm}}$, so that the nulling point still exists at the external magnetic bias to let $\partial_\delta^4 \mathcal{E}_{\textrm{arm}} = 0$.

However, as we increase the stray inductance $\alpha$, which effectively decrease the inductance ratio $\beta$, it causes the ground state to be more and more unstable. Adding to it, increasing $\alpha$ causes the nulling point to shift from $\varphi_\textrm{ext} = 2\pi$ to higher magnetic bias. At a relative large $\alpha$, the nulling point may end up in the unstable regime and become unreachable in real experiment. In Fig.~\ref{fig:alpha_phase}, we plot the ground state stability diagram as we change external magnetic flux and $\alpha$, we further plot shifting of the nulling points as we change $\alpha$ [green curve in Fig~\ref{fig:alpha_phase}]. In Fig.~\ref{fig:alpha_phase}, we set the JRM inductance ratio $\beta = 4.0$ and when $\alpha \sim 0.4$, the nulling point hits the boundary of the unstable regime, which means the nulling point does not exist in experiment any longer.

\bibliography{classical_single_g}

\end{document}